\documentclass{jfm}
\usepackage{graphicx}
\usepackage{epstopdf, epsfig}
\usepackage{float}
\usepackage{subfigure}
\usepackage{amsmath,amssymb}
\usepackage{natbib}
\usepackage{xcolor}
\usepackage{dcolumn}%
\usepackage{bm}%
\newcommand{\lra}[1]{\langle #1 \rangle }

\def\be{\begin{equation}}
\def\ee{\end{equation}}

\shorttitle{Energy fluxes in 3D turbulent Rayleigh-B\'enard convection}
\shortauthor{V. Valori, A. Innocenti, B. Dubrulle and S. Chibbaro}
\title{Weak formulation  and scaling properties of energy fluxes in 3D numerical turbulent Rayleigh-B\'enard convection} 

\author{Valentina Valori\aff{1},
  Alessio Innocenti\aff{2},
  B\'ereng\`ere Dubrulle\aff{1}
 \and Sergio Chibbaro\aff{2}
 \corresp{\email{chibbaro@ida.upmc.fr}}}

\affiliation{\aff{1} SPEC, CEA, CNRS, Universit\'e Paris-Saclay, CEA Saclay, Gif-sur-Yvette, France
\aff{2}Sorbonne Universit\'{e}, CNRS, UMR 7190, Institut Jean Le Rond d'Alembert, F-75005 Paris, France}

\begin{document}
\maketitle

\begin{abstract}
We apply the weak formalism on the Boussinesq equations, to characterize scaling properties of the mean and the standard deviation of the potential, kinetic and viscous energy flux in very high resolution numerical simulations.
The local Bolgiano-Oboukhov length $L_{BO}$ is investigated and it is found that its value may change of an order of magnitude through the domain, in agreement with previous results. 
We investigate the scale-by-scale averaged terms of the weak equations, which are a generalization of the Karman-Howarth-Monin and Yaglom equations.
{ We have not found the classical Bolgiano-Oboukhov picture, but evidence of a mixture of Bolgiano-Oboukhov and  Kolmogorov scalings.
In particular, all the terms are compatible with a Bolgiano-Oboukhov local H\"older exponent for the temperature and
a Kolmogorov 41 for the velocity.}
This behavior may be related to anisotropy and to the strong heterogeneity of the convective flow, reflected in the wide distribution of Bolgiano-Oboukhov local scales. 
The scale-by-scale analysis allows us also to compare the theoretical Bolgiano-Oboukhov length $L_{BO}$ computed from its definition with that empirically extracted through scalings obtained from weak analysis.
{ The key result of the work is to show that the analysis of local weak formulation of the problem is powerful to characterize the fluctuation properties.}
\end{abstract}

\section{Introduction}
The dynamics of flows in natural systems is basically governed by exchanges between  energies of different origin: kinetic, thermal, potential, magnetic, chemical... In the case of oceanic and atmospheric flows, the main energies involved are the kinetic energy and the potential  energy~\citep{vallis2017atmospheric}. Understanding the interplay between the energy fluxes of these quantities, and their scaling properties is one of the main issue for climate modeling. The paradigm to study these issues is the Rayleigh-B\'{e}nard system in which solutions of the Boussinesq  equations describe the movements of velocity $u$ and temperature $T$ of a fluid heated from below. 
{ Even restricting to this configuration, many open issues remain to be addressed~\citep{ahlers2009heat,lohse2010small,chilla2012new}.
We shall focus here on some properties of small scales, describing a new approach to analyse scaling behaviour and more generally fluctuation properties.} 

In realistic conditions, the energy is injected in such systems at large scales under the shape of potential  energy, and converted into kinetic energy and cascaded towards smaller scales by nonlinear interactions. 
{ At scales smaller than the global inhomogeneity scale,} such process can be regarded to occur via globally self-similar processes, with scaling laws that depend upon the parameters and the scale~\citep{Mon_75}. 
In this problem, several characteristic scales have been identified
{ other than the large scale typical of stratification : i) the  kinetic viscous or Kolmogorov scale $\eta$; this scale corresponds to the scale at which the local Reynolds number is of order one. 
Below such scale, statistics of the velocity field are smooth, that is the velocity increments $\delta u_\ell=u(x+\ell)-u(x)$ scale statistically like $\ell$; ii)  the thermal viscous, or Batchelor scale $\eta_T$, corresponding in the same way to the scale at which thermal diffusion becomes dominant. 
Below such scale, also the statistics of temperature field are smooth, that is temperature increments $\delta T_\ell=T(x+\ell)-T(x)$ scale statistically like $\ell$;} iii)  the Bolgiano $L_{BO}$ scale, which is the scale at which buoyancy effects become important and they may balance with dissipative terms. Below such scale, the temperature is usually considered as "passive", with negligible influence on the velocity field. Empirically, the sign, magnitude and scaling of the energy fluxes depend on how these scales are interlinked. 
 
According to the prediction based on a generalization  of the Kolmogorov theory for turbulent fluids, first suggested for stably stratified flows \citep{bolgiano1959turbulent,obukhov1959effect},   the velocity increments and temperature increments scale like $\delta u_\ell\sim \ell^{1/5},\; \delta T_\ell \sim \ell^{3/5}$ above $L_{BO}$,  resulting in a constant flux of potential energy $\partial_\ell \lra{\delta u_\ell (\delta T_\ell)^2}$ towards small scale. In contrast, for scale $\ell<L_{BO}$, 
the kinetic energy flux $\partial_\ell \lra{\delta u_\ell (\delta u_\ell)^2}$ is constant, so that $\delta u_\ell\sim \ell^{1/3}$. 
Unfortunately, several issues make difficult the measurement of the Bolgiano-Oboukhov (BO) scaling in a closed domain~\citep{lohse2010small}. 
In particular, the Bolgiano-Oboukhov length has been found to be globally of the order of the entire volume of the box, and the anisotropy could make also ambiguous to discern between the 
Bolgiano-Oboukhov scaling and other shear-scaling \citep{biferale2005anisotropy}.
Moreover, the similarity argument does not take into account intermittency effect, produced by large fluctuations of velocity gradients or temperature gradients.
 Indeed, it is well known that both velocity and temperature are highly intermittent random fields \citep{benzi1994scaling,cioni1995temperature,lohse2010small}, and therefore local dynamics or local energy  exchange  may be subject to intense fluctuations and strong inhomogeneity. 
 
{ Still it is important to observe that the system is non-homogeneous, because of the presence of the top-bottom and lateral walls.
That makes the Bolgiano-Oboukhov length also a non-homogeneous quantity, which has been shown to vary its value over about one order of magnitude depending on the distance from the walls, both the top-bottom and the lateral ones~\citep{calzavarini2002evidences,kunnen2008numerical,kaczorowski2013turbulent}}
For this reason, the presence of a Bolgiano-Oboukhov scaling seems plausible, possibly over a tiny range.
{  An important issue, common to other non-homogeneous flows, is the difficulty to compute accurate scaling.  
The goal of the present work is deal with this issue presenting a new approach that allows to access local scaling.

Generally speaking, valuable information about the cascade ranges and the scaling-laws of turbulent flows are found by measuring the spectrum of a flow. 
The same content is brought by the second-order structure function, notably employed in numerical simulations. Higher-order structure functions convey more refined information, notably about intermittency~\citep{Fri_95}.
Those are the tools also used to analyze small-scale behavior in Rayleigh-B\'enard convection~\citep{lohse2010small}.
On one hand, structure functions are much noisy, even at the lowest level of second-order, and it is difficult to obtain a clear scaling~\citep{benzi1998heat,calzavarini2002evidences,kaczorowski2013turbulent}.
It is possible to improve the predictions using a particular fitting procedure known as ESS~\citep{Ben_93}. In this case, the ratio between the scaling exponent of different-order structure function can be accessed more neatly. While this procedure has been effective in homogeneous isotropic turbulence where the third-order structure function is known analytically~\citep{Fri_95}, it gives only relative exponents in non-homogeneous flows like RB. 
On the other hand, the most common method to measure the spectrum is using Fourier transforms, 
notably in experiments, and Fourier techniques are inherently global in space and cannot characterize the flow properties locally. Yet the local properties seems to play a crucial role in non-homogeneous flows. 
The purpose of the present work is to put forward an approach capable 
to capture the full complexity of local energy flux and exchanges.
That is relevant for non-homogeneous flows like the Rayleigh-B\'enard convection for the inherent multiscale character of such a flow, as the production, transport and dissipation of energy
depend on the the position in space and on the scale considered.

As discussed in \cite{DubrulleJFM}, a suitable framework is the  weak formulation of the basic equations, via appropriate wavelet transforms. 
Technically, a main advantage wavelets have over Fourier analysis is the identification of flow properties simultaneously as a function of scale and space. For this reason wavelets have already provided their utility to access local scaling in turbulent flows~\citep{meneveau1991dual,farge1992wavelet,jaffard2001wavelets} and most notably the multifractal spectrum~\citep{kestener2004generalizing}.
More generally, the weak approach is related the coarse-grained or filtered equations, which naturally allows to generalise to non-homogeneous flows the scale-by-scale analysis developed for homogeneous turbulence~\citep{duchon2000inertial,Eyink:2006p1379}.
}
In a recent paper~\citep{faranda2018computation}, we have implemented such a framework on the { stable-stratified Boussinesq equations, and derived the generalization of potential, kinetic and viscous energy flux in scale.
A preliminary yet encouraging analysis has then been made of some atmospheric data. 
By construction, the atmospheric data only provide} a fragmented view of the energy fluxes, because they do not extend all the way to the viscous scales, where the extremes of  kinetic energy flux are found to happen \citep{Saw2016}. 
In the present paper, we therefore apply the analysis to a more controlled and cleaner system, provided by  numerical solutions of Boussinesq equations at high resolution. 

The paper is organized as follows: In Section II, we first recall the theoretical model and the numerical method. In Section III, we describe the theoretical framework and tools. In section IV, we provide the results and we discuss them. Finally we draw conclusions.

\section{Governing equations and numerical method}

We consider a turbulent Rayleigh-B\'enard convection, in which a horizontal fluid layer is heated from below. 
Horizontals and wall-normal coordinates are indicated by $x$, $y$ and $z$, respectively.
Using the Boussinesq approximation, the system is described by the following dimensionless balance equations
\begin{eqnarray}
\frac{\partial u_i}{\partial x_i}&=&0, \\
\label{c1}
\frac{\partial u_i}{\partial t} +  u_j\frac{\partial u_i}{\partial x_j} &=& -\frac{\partial P}{\partial x_i} + \sqrt{\frac{Pr}{{Ra}}}\frac{\partial^2 u_i}{\partial x_j^2}+\delta_{i,3} \theta, \\
\label{ns1}
\frac{\partial \theta}{\partial t} +  u_j\frac{\partial \theta}{\partial x_j} &=&  \frac{1}{\sqrt{ Pr Ra}}\frac{\partial^2 \theta}{\partial x_j^2}, 
\label{en1}
\end{eqnarray}
where $u_i$ is the $i^{th}$ component of the velocity vector, $P$ is pressure, $\theta= (T-T_{0})/\Delta T$ is the dimensionless temperature, $\Delta T= T_{H} - T_{C} $ is the imposed  temperature difference  between the hot bottom wall ($T_H$) and top cold wall ($T_C$), $T_0=(T_C+T_H)/2$, whereas $\delta_{1,3} \theta$ is the driving buoyancy force  (acting in the vertical direction $z$ only).
The reference velocity is $u_{ref}=(g \beta H \Delta T  )^{1/2}$, with $H$ the domain height, $g$ the acceleration due to gravity and $\beta$ the thermal expansion coefficient.
The Prandtl and the Rayleigh numbers in Eqs. (\ref{c1})-(\ref{en1}) are defined as $Pr=\nu/\kappa$ and $Ra= (g \beta \Delta T H^3) / (\nu k)$, with  $\nu$ the fluid kinematic viscosity and $\kappa$ the thermal diffusivity.

Two DNS are performed for a fixed $Pr=1$ at $Ra=10^7$ and $10^8$ in a cubic box of size $H^3$ with the $x-y$ plane parallel to the horizontal plates and the z axis pointing in the direction opposite to that of gravitational acceleration.
For the velocity, no-slip boundary conditions are used everywhere, like in experiments.
For the temperature instead,  adiabatic conditions are imposed in all lateral sidewalls whereas isothermal conditions are used on the top and bottom plates. 
Table \ref{tab1} reports the main parameters of the simulations. 
{ Previous studies indicate that BO scaling might more clearly observable at the moderate Ra numbers chosen here~\citep{lohse2010small}. 
In order to make the scaling neater, it would be 
in principle helpful to use higher Prandtl numbers~\citep{kaczorowski2013turbulent}.
However, the dependence is very slow and the computational effort needed to keep the present accuracy at much higher Pr numbers very important.
For this first study concerning the weak approach to turbulent convection we have thus preferred use the somewhat standard  $Pr=1$. The weak analysis of a higher Pr case is planned for a future study.
}

Equations (\ref{c1})-(\ref{en1}) are solved through the open-source code Basilisk (see http://www.basilisk.fr/).
In particular, space is discretized using a Cartesian (multi-level or tree-based) grid where the variables are located at the center of each control volume (a square in 2-D, a cube in 3-D) and at the center of each control surface.
Second-order finite-volume numerical schemes for the spatial gradients are used \citep{popinet2003gerris,popinet2009accurate,lagree2011granular}.
Navier-Stokes equations are integrated by a projection method \citep{chorin1969convergence}, and
the time advancing is made through a fractional-step method using a staggered discretization in time of the velocity and the scalar fields \citep{popinet2009accurate}.

Some remarks are in order concerning the numerical method.
From a numerical point of view, it is worth noting that our upwind method is actually third-order in space when considering the advection term. 
The Basilisk code has been tested in isotropic turbulence and in particular compared with a finite-volume scheme which preserves energy \citep{fuster2013energy}. 
The results are in good agreement with those obtained with a spectral code and no difference is encountered between the two volume-finite methods, { whenever the resolution is sufficient to resolve  all scales} \footnote{http://www.basilisk.fr/}.
Furthermore, the code has been recently used and extensively validated in Rayleigh-B\'enard turbulent convection \citep{castillo2016reversal,castillo2017turbulent}.
In that thorough validation it has been shown that, { provided the requirements on the resolution are respected~\citep{stevens2010radial,shishkina2010boundary}}, no appreciable difference can be found with respect to the literature with respect to any observable.
Furthermore, in a recent work the numerical approach has been assessed in a one-to-one comparison against a standard code also in a case of atmospheric boundary layer. The results are satisfying in all respects and numerical dissipation appears to be ineffective, provided the resolution is sufficient to well resolve the boundary layer~\citep{van2017adaptive}.
\begin{table}
\begin{center}
\begin{tabular}{lccccccccccc}

 Case & $Ra$ & $Pr$ & $N_x \times N_y \times N_z $ & $\frac{\Delta}{\eta_{bulk}}$  & $\frac{\Delta}{\eta_{BL}}$ & $\Delta t$ & $N_T$ &
 $Nu$ &  $Nu_{\epsilon_u}$ & $Nu_{\epsilon_T}$  \\[4pt]

 A &$10^7$ & $1.0$ &$1024 \times 1024 \times 1024$ & 1/10 & 1/8 & 0.0015 & 3.3/30 &
 15.8 & 16 & 15.9 \\

 B &$10^8$ & $1.0$ & $1024 \times 1024 \times 1024$ & 1/8 & 1/4 &  0.001 & 4.2/15
 & 31.1 & 31.3 & 31.8 \\
 
\end{tabular}

\caption {List of the dimensionless parameters, $Ra$ and $Pr$, for the different test runs, and the parameters of the simulations: The   number   of   grid   points
$ N_x \times N_y \times N_z$   in   the   respective   spatial   direction; the number of grid points required for resolving the thermal boundary layer $N_T$ { (requirement/actual resolution); The requirement is based on the analysis in~\citet{stevens2010radial,shishkina2010boundary}}.
The mean heat transfer computed with the three different formulas:
$Nu\equiv1+\sqrt{Ra Pr} \lra{u_z T}~,~Nu_{\epsilon_u}\equiv 1+\sqrt{Ra Pr}\lra{\epsilon_u}~\mathrm{and}~ Nu_{\epsilon_T}\equiv \sqrt{Ra Pr} \lra{\epsilon_T}$
{ , where $ \lra{}$ indicate averaging, and the statistics have been computed averaging in space over the entire volume and over $300$ reference time.}} \label{tab1} 
 \end{center}
\end{table}

We have chosen a horizontal ($N_x , N_y$) and vertical ($N_z$) number of points sufficiently high to solve the smallest length scale of the problem, which is the Kolmogorov length scale $\eta=\left(\nu^3/\lra{\epsilon}\right)^{1/4}$, since $Pr = 1$ in all regions. 
Moreover, for Rayleigh-B\'{e}nard convection criteria have been proposed to ensure proper resolution of the thermal dynamics~\citep{shishkina2010boundary}.
Although this particular resolution is only required near to the wall, we have chosen to use an uniform grid since we are interested in monitoring the fluctuations in the center of the cube, where vertical non-homogeneity is less important. 
For this reason and given the geometry chosen, the grid spacing is the same in all directions $\Delta_x=\Delta_y=\Delta_z=\Delta=1/1024$. 
The Kolmogorov length scale $\eta$  is computed a posteriori from the datasets via spatial and time average. 
As shown in table 1, for both the DNS, the value of the Kolmogorov length scale is much larger than the grid spacing, so that we are over-resolving the flow~ \citep{verzicco2003numerical}.
Moreover as shown in the same table 1, the resolution greatly exceeds also the requirements about the thermal layers, notably at lower Prandtl.
That was a deliberate choice for two reasons:
(i) Basilisk is a volume-finite code which may add some numerical diffusion at the smallest scales. Since we are precisely interested in the behaviour at small scales, we use a resolution higher than necessary to avoid spurious effects.
(ii) We are also interested in the possible presence of extreme events at very small scales, $\ell \sim \eta$, and therefore we have carried out simulations with a resolution much higher than usual to be sure to well resolve all the scales $ \ell \lesssim \eta$.

After the initial transient, velocity and temperature fields, are collected with a time interval significantly longer than the large eddy turnover time $2h/u_{ref}$ in order to assure that the fields  are uncorrelated.
The statistical convergence has been checked looking at different statistics and observing that the average mean field is zero.
In order to assess the resolution of the numerical method, we show in table 1 also the consistency relation for the mean heat transfer \citep{siggia1994high,verzicco2003numerical}: $Nu\equiv1+\sqrt{Ra Pr} \lra{u_3 \theta}=Nu_{\epsilon_u}\equiv 1+\sqrt{Ra Pr} \lra{\epsilon_u}= Nu_{\epsilon_T}\equiv \sqrt{Ra Pr} \lra{\epsilon_T}$, where $ \lra{}$ indicate ensemble averaging, $\epsilon_u$ is the dissipation-rate and $\epsilon_T$ is the temperature-variance dissipation rate.
The results are indeed consistent and in agreement with the values from the literature for both $Ra$ numbers~\citep{ahlers2009heat}.

\section{Weak formulation}
\subsection{Summary of local energy budget}
{ In this section, we present the local energy budget of Boussinesq equations (\ref{c1})-(\ref{en1}), which relies on weak formulation of the equations~\citep{duchon2000inertial} and has been  derived in a recent work \citep{faranda2018computation} for the case of stable-stratified flows.
Although we deal with unstable stratification in this work, the derivation is the same and we refer to that reference for the details.
The local budget involves  filtered or coarse-grained observable $\widetilde{o}_{\ell}$  defined as
\begin{equation}
\widetilde{o}_{\ell}(x,t) \equiv \int_{}^{} d^d r G_{\ell}(r) {o}(x+r,t) ~,
\end{equation}
where the  subscript $\ell$ refers to the scale   dependence introduced by the filtering.
The filter $G$ is a smooth  function, non-negative, spatially localized and such that $\int d\vec r \ G(\vec r)=1$, and $\int d\vec r \ \vert \vec r \vert ^2 G(\vec r) \approx 1$. The function $G_\ell$ is rescaled with $\ell$ as $G_\ell (\vec r) = \ell^{-3}G(\vec r/\ell)$. At a given finite scale $\ell$, the local energy budget depends on this filtering, but all the results obtained in the limit $\ell\to 0$ are independent of $G$. In the sequel, we take $G$ as a Gaussian whenever analyzing the numerical data.}
The local scale-by-scale equations for the energy and temperature-variance read as \citep{faranda2018computation}:
\begin{eqnarray}\label{filter-u}
&\partial_t&  (\frac{1}{2}  \textbf{u}\cdot \widetilde{\textbf{u}}_\ell )+ \nabla \cdot \Bigg[\frac{1}{2}(\textbf{u} \cdot \widetilde{\textbf{u}}_{\ell})\textbf{u} + \frac{1}{2}(p \widetilde{\textbf{u}}_{\ell}+\widetilde{p}_{\ell}\textbf{u}) + \frac{1}{4} \widetilde{(\mid \textbf{u} \mid^2 \textbf{u})}_{\ell} \nonumber \\
& & - \frac{1}{4} \widetilde{(\mid \textbf{u} \mid^2)}_{\ell} \textbf{u} 
- \frac{1}{2}{\sqrt{\frac{Pr}{Ra}}}\nabla (\widetilde{\textbf{u}_\ell} \cdot \textbf{u})
\Bigg] = -\frac{1}{4} \int_{}^{} d^dr \nabla G_{\ell} \cdot \delta{\textbf{u}}(r) \mid \delta \textbf{u}(r) \mid^2, \nonumber\\
& & -  {\sqrt{\frac{Pr}{Ra}}}
  \int_{}^{} d^dr \nabla^2 G_{\ell} \mid\delta \textbf{u}(r) \mid^2 
 + \frac{1}{2}  (u_3\widetilde{\theta}_{\ell} + \widetilde{u}_{3\ell}\theta) \\ 
&\equiv& -\textrm{D}_{\ell}-\textrm{D}^{\nu}_{\ell}+\textrm{D}^c_{\ell}~, \nonumber
\end{eqnarray}
and
\begin{eqnarray}\label{filter-t}
&\partial_t &(\frac{1}{2} \theta\widetilde{\theta}_{\ell}) + \nabla \cdot \Bigg[\frac{1}{2}(\textbf{u} \cdot \widetilde{\theta}_{\ell}) \theta + \frac{1}{4} \widetilde{( \theta^2 \textbf{u})} _{\ell} -\frac{1}{4} \widetilde{(\theta^2)}_{\ell} \textbf{u})) 
 - \frac{1}{2 \sqrt{Ra Pr}}  \nabla ( \theta \widetilde{\theta_\ell})
 \Bigg]  \nonumber
 \nonumber \\
&=& -\frac{1}{4} \int_{}^{} d^dr \nabla G_\ell\cdot \delta{\textbf{u}}(r) (\delta \theta)^2 - \frac{1}{\sqrt{Pr Ra}}   \int_{}^{} d^dr \nabla^2{G}_{\ell} \mid\delta \theta(r) \mid^2,  \\ 
&\equiv& -\textrm{D}^T_{\ell}-\textrm{D}^{\kappa}_{\ell}~. \nonumber
\end{eqnarray}

{ 
where 
$\delta \theta(\textbf{r})=\theta(\textbf{x+r})-\theta(\bf{x})$ is the temperature increment over a vector $\textbf{r}$, and similar definition for $\delta \bf{u}(\textbf{r})$.
All the budget terms denoted by $\textrm{D}$ are implicitly defined by the equations, as indicated by the symbol $\equiv$.
In particular, in eq. (\ref{filter-u}) $\textrm{D}_{\ell}$ is related to the inertial dissipation, $\textrm{D}^{\nu}_{\ell}$ represents the viscous dissipation and $\textrm{D}^c_{\ell}$ the coupling term.
In eq. (\ref{filter-t}), $\textrm{D}^T_{\ell}$
is the inertial dissipation term for the temperature variance and $\textrm{D}^{\kappa}_{\ell}$ represents the viscous diffusion.
}
As discussed in \cite{faranda2018computation}, \cite{DubrulleJFM}, and shown below, these equation are a local fluctuating form of the K\'arm\'an-Howarth-Monin (KHM) equations~\citep{Mon_75}, including the exchange term between temperature and velocity due to buoyancy.
{ As typical for non-equilibrium macroscopic phenomena, the  kinetic or thermal  energy $\partial_t  (\frac{1}{2}  \textbf{u}\cdot \textbf{u}_\ell )$ or $\partial_t (\frac{1}{2} \theta\widetilde{\theta}_{\ell}) $ evolves through (i) a current describing mean  transport  via a  spatial flux; (ii) a local term related to the exchange of energy at the scale $\ell$, (iii) a local sink term due to (viscous or thermal) dissipation, and (iv) a local term linked to the buoyancy work that redistributes the energy between the 
thermal and the kinetic part.

As conjectured by Onsager \citep{eyink2006onsager} and rigorously stated by \cite{duchon2000inertial}, the non-linear inter-scale terms $\textrm{D}_{\ell}$ converges to the inertial dissipation term at infinite Reynolds number in the asymptotic small-scale limit 
\begin{equation}
\textrm{D} = \lim_{ \ell \rightarrow 0} (\lim_{\nu\rightarrow 0} \textrm{D}_{\ell})~.
\end{equation}
In this limit, the nonlinear terms may  dissipate energy, when the field is sufficiently irregular \citep{duchon2000inertial}.
In the same way, we define
\begin{equation}
\textrm{D}^T = \lim_{ \ell \rightarrow 0}  \textrm{D}_{\ell}^T~,~
\textrm{D}^\nu = \lim_{ \ell \rightarrow 0}  \textrm{D}_{\ell}^\nu~,~
\textrm{D}^\kappa = \lim_{ \ell \rightarrow 0}  \textrm{D}_{\ell}^\kappa~,~
\textrm{D}^c = \lim_{ \ell \rightarrow 0}  \textrm{D}_{\ell}^c,
\end{equation}

Furthermore, in the average sense we have the  relation
\begin{equation}
\lra{\textrm{D}^{\nu}}\equiv \mathcal{D}^{\nu} = {\epsilon}~,~~\text{with}~~ \epsilon=\lra{\nu \vert \nabla {u} \vert^2}~,
\end{equation}
and 
\begin{equation}
\lra{\textrm{D}^{\kappa}} \equiv \mathcal{D}^{\kappa} = {\epsilon_T}~,~~\text{with}~~ \epsilon_T=\lra{\kappa \vert \nabla {\theta} \vert^2}~;
\end{equation}
where $\lra{}$ denotes spatial and time average.
In the same way we have
\begin{equation}
\lra{\textrm{D}^c}\equiv \mathcal{D}^c= (Nu-1)/\sqrt{Ra Pr}~,~~\text{with}~~ (Nu-1)/\sqrt{Ra Pr}=\lra{u_3\theta}~.
\end{equation}
 On the other hand, whenever $u$ and $\theta$ are regular, we have in the limit $\ell\to 0$, $\delta u\sim \ell$ and $\delta \theta\sim \ell$, 
so that both $\textrm{D}$ and $\textrm{D}^T$ scale like $\ell^2$ and tends to zero. The location where these quantities do not converge with zero is the location of potential 
quasi-singularities \citep{DubrulleJFM} that will be studied elsewhere. Empirically, we observe that these points are very rare, so that on average, we have $\lra{\textrm{D}}=\lra{\textrm{D}}^T=0$ .
The way how these limits are achieved is however informative about the scaling properties of the flow, as is shown in Section \ref{sec:KO}.
When averaged, equations (\ref{filter-u})-(\ref{filter-t}) give the general forms of the mean energy and temperature budgets, and
they are interesting since they provide a scale-by-scale way to analyse turbulent flows, as highlighted in several recent works focused on anisotropic turbulent flows~\citep{hill1997applicability,danaila1999generalization,rincon2006anisotropy,cimarelli2013paths,gauding2014generalised,togni2015physical,mollicone2018turbulence}.
}

\subsection{Global energy budget and Yaglom equations}
{ By taking ensemble averages of the equations (\ref{filter-u}) and (\ref{filter-t}), we can also obtain a global scale-dependent energy budget. 
Considering a stationary state, and taking into account the contribution of the spatial flux terms due to the temperature boundary condition, we then obtain:
\begin{eqnarray}
\frac{1}{2}  \lra{\textbf{u}\widetilde{\theta}_{\ell} + \widetilde{\textbf{u}}_{\ell}\theta} &=&\frac{1}{4} \int_{}^{} d^dr \nabla G_{\ell} \cdot \lra{\delta{\textbf{u}}(r) \vert \delta \textbf{u}(r)\vert^2} 
+
 \sqrt{\frac{Pr}{Ra}}
  \int_{}^{} d^dr \nabla^2 G_{\ell} \lra{\delta( \textbf{u}(r))^2 },\nonumber\\
 \oint_{\partial V} J_\ell^Td\Sigma &=&\frac{1}{4} \int_{}^{} d^dr \nabla G_\ell\cdot \lra{\delta{\textbf{u}}(r) (\delta \theta)^2}
+\frac{1}{\sqrt{Pr Ra}}  \int_{}^{} d^dr \nabla^2{G}_{\ell} \lra{(\delta \theta(r))^2} ,
\label{almostYaglom}
\end{eqnarray}
where $J_\ell^T=\left[\frac{1}{2}(\textbf{u} \cdot \widetilde{\theta}_{\ell}) \theta + \frac{1}{4} \widetilde{( \theta^2 \textbf{u})} _{\ell} -\frac{1}{4} \widetilde{(\theta^2)}_{\ell} \textbf{u})) 
 - \frac{1}{2 \sqrt{Ra Pr}}  \nabla ( \theta \widetilde{\theta_\ell})\right]$.
 It is worth noting that in the present work 
statistical averages will be computed through spatial and time averaging, thanks to the stationarity of the flow and by using the Ergodic hypohtesis.
In the limit $\ell\to 0$, we have $\oint J_\ell^Td\Sigma \to-Nu/\sqrt{Ra Pr}$ so that the global energy budget yields:
\begin{eqnarray}
(Nu-1)/\sqrt{Ra Pr}&=&\lra{\textrm{D}}+ \epsilon,\nonumber\\
 Nu/\sqrt{Ra Pr}&=&\lra{\textrm{D}}^T +\epsilon_T.
 \label{schematicss}
 \end{eqnarray}
Taking into account $\lra{\textrm{D}}=\lra{\textrm{D}}^T=0$, we then get $\epsilon=(Nu-1)/\sqrt{Ra Pr}$ and $\epsilon_T=Nu/\sqrt{Ra Pr}$
which are the non-dimensional global energy budget equations for Rayleigh-B\'enard, first derived by \cite{siggia1994high}.
For a finite scale, the global energy budget  Eq. (\ref{almostYaglom}) reads schematically
\begin{eqnarray}
{\cal{D}}^c_{\ell}&=&\cal{D}_{\ell}+\cal{D}^{\nu}_{\ell},\\ \nonumber
\oint_{\partial V} J_\ell^Td\Sigma
&=&{\cal{D}}_{\ell}^{T}+\cal{D}^{\kappa}_{\ell},
 \label{schematic}
 \end{eqnarray}
and describes energy cascades through scale for both temperature and velocity.}
{ \subsection{Special Length scales}\label{sec:GC}
The global budgets  Eq. (\ref{schematic})  provides systematic definitions of characteristics scales that traces the boundary between diffusive and inertial behaviour.
 There are indeed two interesting scales corresponding to situations where:
 \par $\bullet\; \cal{D}_{\ell}=\cal{D}^{\nu}_{\ell}$; the corresponding scale is $\eta$, the dissipative scale \citep{DubrulleJFM}.
  \par $\bullet\; {\cal{D}}_{\ell}^T=\cal{D}^{\kappa}_{\ell}$; the corresponding scale is $\eta_T$, the thermal dissipative scale.
 Further, one could also define the scale at which 
 \par $\bullet\; \cal{D}_{\ell}=\cal{D}^{C}_{\ell}$; the corresponding scale would then correspond to a global Bolgiano scale.
 On a practical side, however, these definition are not easy to handle, since they are true only after averaging over the whole domain. In the sequel, we shall then rely on definitions involving $\epsilon$ and $\epsilon_T$.}

\subsection{General scalings}\label{sec:KO}
Let us now make a general theoretical analysis.
Consistently with the Kolmogorov-Onsager framework~\citep{paladin1987anomalous,Fri_95,eyink2006onsager}, 
we assume that velocity and temperature increments are globally H\"older continuous fields with exponent $h$:
\begin{equation}
\vert \delta {\bf u}({\bf x,r})\vert
 \sim r^{h^u}~~;~~
 \vert\delta { \theta}({\bf x,r}) \vert
 \sim r^{h^T},
 \label{eq:multifra}
\end{equation}
where velocity exponent is denoted by $h^u$ and the temperature one by $h^T$. 
We do not consider here the \emph{local} properties, which are related to anomalous scaling and large deviations~\citep{benzi1984multifractal,paladin1987anomalous,boffetta2008twenty}.

If  our system is \emph{locally isotropic},  everything depends only on the module of the difference in position $r=\vert {\bf r}\vert$ and the scaling exponent for the velocity is the same for the horizontal and vertical component. We shall see in sequel that such hypothesis is probably not satisfied.
Indeed, in such a case from the scaling of   $\delta u_\ell^{h^u}$ and 
$\delta \theta_\ell^{h^T}$ one can deduce that:
\begin{equation}
\mathcal{D}_{\ell}  \sim  \ell^{3h^u-1}, ~
\mathcal{D}^{T}_{\ell} \sim  \ell^{h^u+2h ^T-1} , ~
\mathcal{D}^c_{\ell} \sim \ell^{h^u+h ^T}, ~
\mathcal{D}^{\nu}_{\ell}\sim \ell^{2h^u-2}, ~
\mathcal{D}^{\kappa}_{\ell}\sim \ell^{2h ^T-2}.
\label{scalingsD}
\end{equation}
In this framework, Kolmogorov scaling gives $h^u=1/3$, and $h^T=1/3$. That should mean 
\begin{equation}
\mathcal{D}_{\ell}\sim \ell^0~, ~\mathcal{D}^T_{\ell} \sim \ell^0~, \mathcal{D}^{\nu}_{\ell} \sim \ell^{-4/3} ,
 \mathcal{D}^{\kappa}_{\ell} \sim \ell^{-4/3}~,~
 \text{and}~ \mathcal{D}^{c}_{\ell} \sim \ell^{2/3}.
\end{equation}
This regime is obtained when, in eq. (\ref{almostYaglom}), the energy transfer is provided both by the thermal and the kinetic component: $\mathcal{D}_{\ell}\sim \epsilon~,~\mathcal{D}^T_{\ell}\sim \epsilon^T$.

Instead, in the Bolgiano-Oboukhov range, the energy transfer is provided by the thermal component, while the inertial term in the kinetic energy is affected by the exchange term. This corresponds to $\mathcal{D}_{\ell}\sim \mathcal{D}^c_{\ell}$ and $\mathcal{D}^T_{\ell}\sim \epsilon_T$, resulting in $h^u=3/5$ and $h^T=1/5$, and therefore
\begin{equation}
\mathcal{D}_{\ell}\sim \ell^{-4/5}~, 
~\mathcal{D}^T_{\ell} \sim \ell^{0}~, 
\mathcal{D}^{\nu}_{\ell} \sim \ell^{-4/5}~,~
 \mathcal{D}^{\kappa}_{\ell} \sim \ell^{-8/5}~,~
 \text{and}~ \mathcal{D}^{c}_{\ell} \sim \ell^{4/5}
\end{equation}
Hence, by looking at scaling properties of these quantities and at their balance, we can infer consistency with K41 or Bolgiano scaling. The scalings are summarized in table \ref{tab:scaling}.
{ It is worth emphasising that these scalings are obtained through similarity arguments based on the locally isotropic hypothesis, so that deviations may be observed whenever this hypothesis is not fulfilled, notably for the coupling term.}

An important empirical observations is that the standard deviation of the observables provides cleaner scaling laws (Section \ref{sec:scaling}) than the average. 
{ We have not yet fully understood this issue, but the same trend has been found in independent experiments~\citep{Saw2016,saw2018universality}.
 We think this is because fluctuations are more robust to changes of the orientation of the large-scale circulation of the flow than the time average.
It is also possible that squared observables scale better because always positive, similarly to what encountered in calculating structure functions ~\citep{benzi1994scaling}.
We cannot however be assertive on this point, since we cannot use absolute value in the filtering terms, and time-averaging the absolute value of different quantities turns out to be rather inconclusive.}
 
 \begin{table}
\begin{center}
\def~{\hphantom{0}}
\begin{tabular}
{cccccc}%
    Quantity&General &Kolmogorov 41 &Bolgiano  & Present study  & Present study\\
    & & & & \emph{Theoretical Conjecture} & \emph{Numerical fit}\\[4pt]
 \hline
  &$h^u$  &  $1/3$ &  $3/5$ &  $1/3$ & $0.34\pm 0.025$\\
  &$h^T$  &  $1/3$ &  $1/5$ &  $1/5$ &$0.18\pm 0.04$ \\
\hline
$\mathcal{D}_{\ell}$& $ \ell^{3h^u-1}$ & $\ell^{0}$ & $\ell^{4/5}$  & $\ell^{0}$  & $\ell^{0.04\pm 0.06}$ \\

$\mathcal{D}^T_{\ell}$& $ \ell^{h^u+2h^T-1}$ & $\ell^{0}$ & $\ell^{0}$  & $\ell^{-4/15}$ & $\ell^{-0.3\pm 0.12}$\\

$\mathcal{D}^\nu_{\ell}$& $ \ell^{2h^u-2}$ & $\ell^{-4/3}$ & $\ell^{-4/5}$  & $\ell^{-4/3}$ & $\ell^{-1.4\pm 0.07}$\\

$\mathcal{D}^\kappa_{\ell}$& $ \ell^{2h^T-2}$ & $\ell^{-4/3}$ & $\ell^{-8/5}$  & $\ell^{-8/5}$ & $\ell^{-1.66\pm 0.06}$\\

$\mathcal{D}^c_{\ell}$& $ \ell^{h^u+h^T}$ & $\ell^{2/3}$ & $\ell^{4/5}$  & $ \slash $ & $ \slash $\\

\end{tabular}
\caption{Summary of scaling laws for the different quantities appearing in equations (\ref{filter-u}) and (\ref{filter-t}), depending upon the scaling of velocity and temperature increments.
{ The last two columns refer to the present study. 
The last one indicates the values extracted via a fit procedure of the data presented in section \ref{sec:scaling}. We have used only $Ra=10^8$ since the scaling range is larger. The exponents obtained by data at $Ra=10^7$ are however consistent with these values.}
}
\label{tab:scaling}
  \end{center}
\end{table}

\section{Results}

\subsection{Flow Fields}
\begin{figure}
\begin{center}
{\includegraphics[width=1\textwidth]{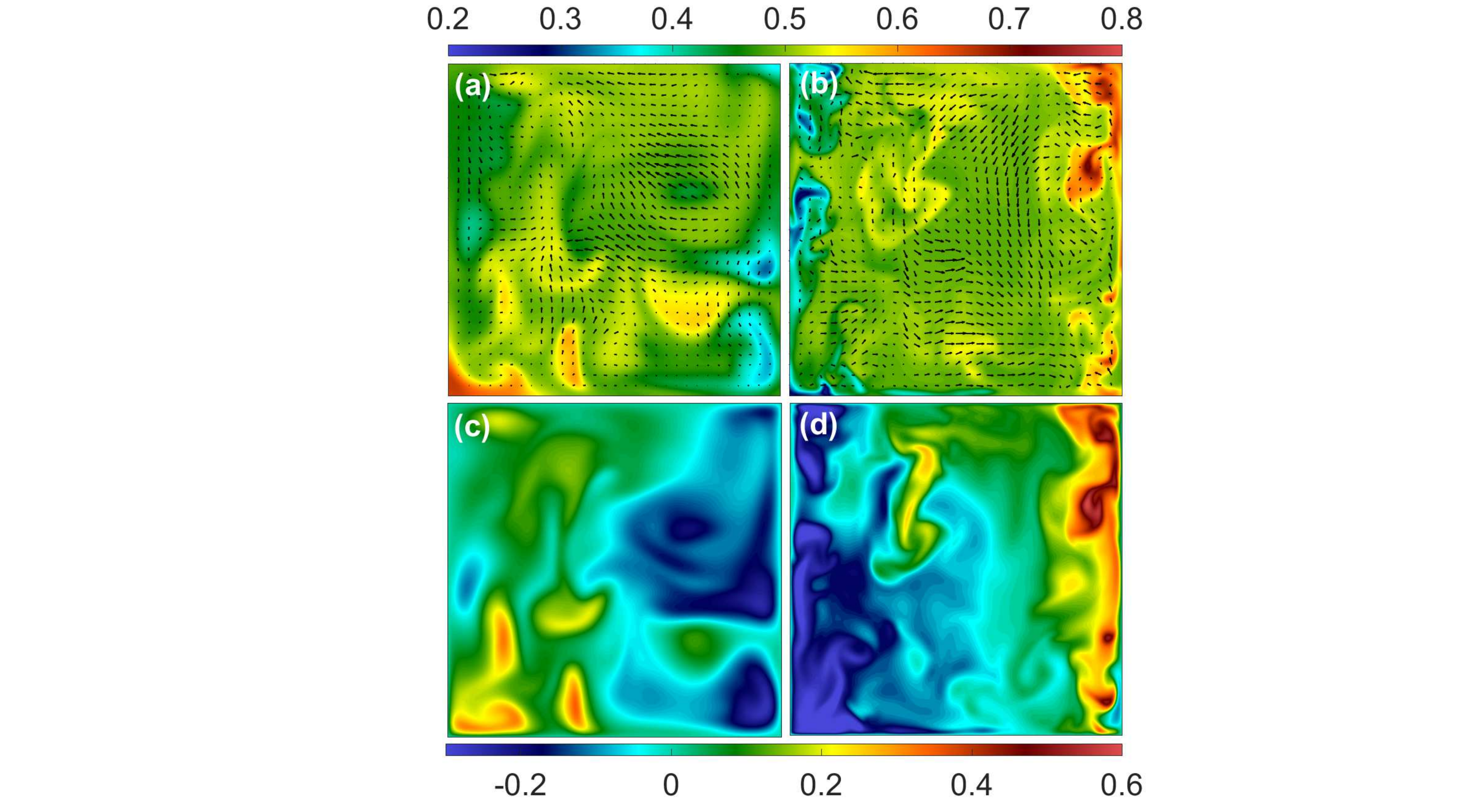}}
\caption{Panel (a) and (b): Instantaneous temperature and velocity fields in the horizontal $x-y$ plane midway between the vertical walls. The velocity is superposed as vectors on the temperature map. {The temperature colorbar is at the top of the figure.} Panel (c) and (d): Instantaneous heat fluxes, at the same instant of time and horizontal cross-section of the plots in panels (a) and (b).  {The heat flux colorbar is at the bottom of the figure.} 
Results for different $Ra$ are displayed, $Ra=10^7$ in panels (a) and (c), and $Ra=10^8$ in panels (b) and (d).}
\label{Fig:1}
\end{center}
\end{figure}	
In figure \ref{Fig:1} we show a horizontal slice of instantaneous temperature and velocity fields at the centre of the domain, region we focus on in this work. 
It seems established that coherent thermal and velocity structures, the so-called thermal plumes, defined as a localized portion of fluid having a temperature contrast with the background, play a major role in the transport of heat in turbulent convection \citep{chilla2012new,shang2003measured}.
These structures emerge from the dynamics of the boundary layer, and in the centre of the cell they are heavily impacted by the large-scale circulation (LSC), and the { geometrical aspect of the flow may be quite different}.
Indeed, in figure \ref{Fig:1} it is seen that cold plumes are directed mostly along one side of the cell while the hot plumes go up at the other side, because of the impingement of the fluid caused by LSC.
These results are in line with experiments made in different geometries \citep{zhou2007morphological,liot2016simultaneous}.

\subsection{Local Bolgiano-Oboukhov length-scale}

{ The visualisation of the fields together with the numerical assessment summarised in table \ref{tab1} ensure that present simulations are accurate.
Before going into the main subject of the present work, which is to use the weak formulation of the equations, it is useful to study the properties of the Bolgiano-Oboukhov scale $L_{BO}$.}

The Bolgiano-Oboukhov length is an estimate of the distance at which the buoyancy and dissipative terms balance in the Boussinesq equations. 
It also represents the scale at which temperature cannot be considered passive anymore and therefore different scalings are expected.
It is defined on a dimensional ground as
\begin{equation}
\hat L_{BO}\equiv (\beta g)^{-3/2}\lra{\hat{\epsilon}}^{5/4}\lra{\hat{\epsilon_T}}^{-3/4}~,
\label{eq:LOB}
\end{equation}
{ where we have used the hat $\hat{}$ symbol to highlight that the quantities in this formula are dimensional.
A local version of this length can be defined as 
\begin{equation}
L_{BO}^{\textrm{local}}(x,y,z)\equiv \lra{\epsilon}_t^{5/4}\lra{\epsilon_T}_t^{-3/4}~,
\label{eq:LOB-norm}
\end{equation}
where we use the non-dimensional dissipations.
In this definition the length is obtained averaging on time but not over the volume.
It is known \citep{benzi1998heat,calzavarini2002evidences} that  $L_{BO}^{\textrm{local}}$ depends on the position in the convection cell and on the boundary conditions.
}
In particular, since we focus on the bulk region, we have computed the average 
\begin{equation}
L_{BO}^{\textrm{bulk}}=\lra{\epsilon^{5/4}}_{x,y,t}
\lra{\epsilon_T}_{x,y,t}^{-3/4}~,
\label{eq:lb-bulk}
\end{equation}
at $z=0$, that is in the middle of the cube.
{The average is made in time and over the bulk region defined as a square of side $0.8H$.
This choice will be explained later in section \ref{sec:wavelet}.}
The Bolgiano-Oboukhov length may also be roughly estimated using the $0$-th law of turbulence and similar scaling for the temperature $\epsilon\sim \frac{u_{rms}^3}{L}$, $\epsilon_T\sim \frac{u_{rms} (\lra{T^{\prime}})^2}{L}$, where we have considered a typical fluctuation of temperature $T^{\prime}$ and a typical length scale $L$ for velocity and temperature.
Considering in the present case that these length scales are of the same order of the typical distance from the plates $z_*$, we get for a global length yet depending on the vertical coordinate:
$\hat L_{BO}=(\beta g)^{-3/2}u_{rms}^3\lra{T^{\prime}}^{-3/2}z_*^{-1/2}$.
It is seen from this expression that in the centre of the cell the estimate is $\hat{L_{BO}}\sim H$, that is of the order of the cell dimension. 
{ The global averaged Bolgiano length turns out to be of the order of the entire height of the cube, indeed.
In particular, we have found  $L_{BO}^{\textrm{bulk}}\lesssim 0.5$ for both $Ra$.}
However, given that our problem is non-homogeneous with walls at the boundaries, the fluctuation statistics are also dependent on the distance from sidewalls and thus this estimate does not permit to access to the local behaviour.
We can only expect that very near to the walls, $L_{BO}$ attains it s maximum value. { In particular, the local length formally diverges at the lateral walls because of the adiabatic conditions.}
\begin{figure}
\begin{center}
\begin{subfigure}[]
{\includegraphics[scale=0.31]{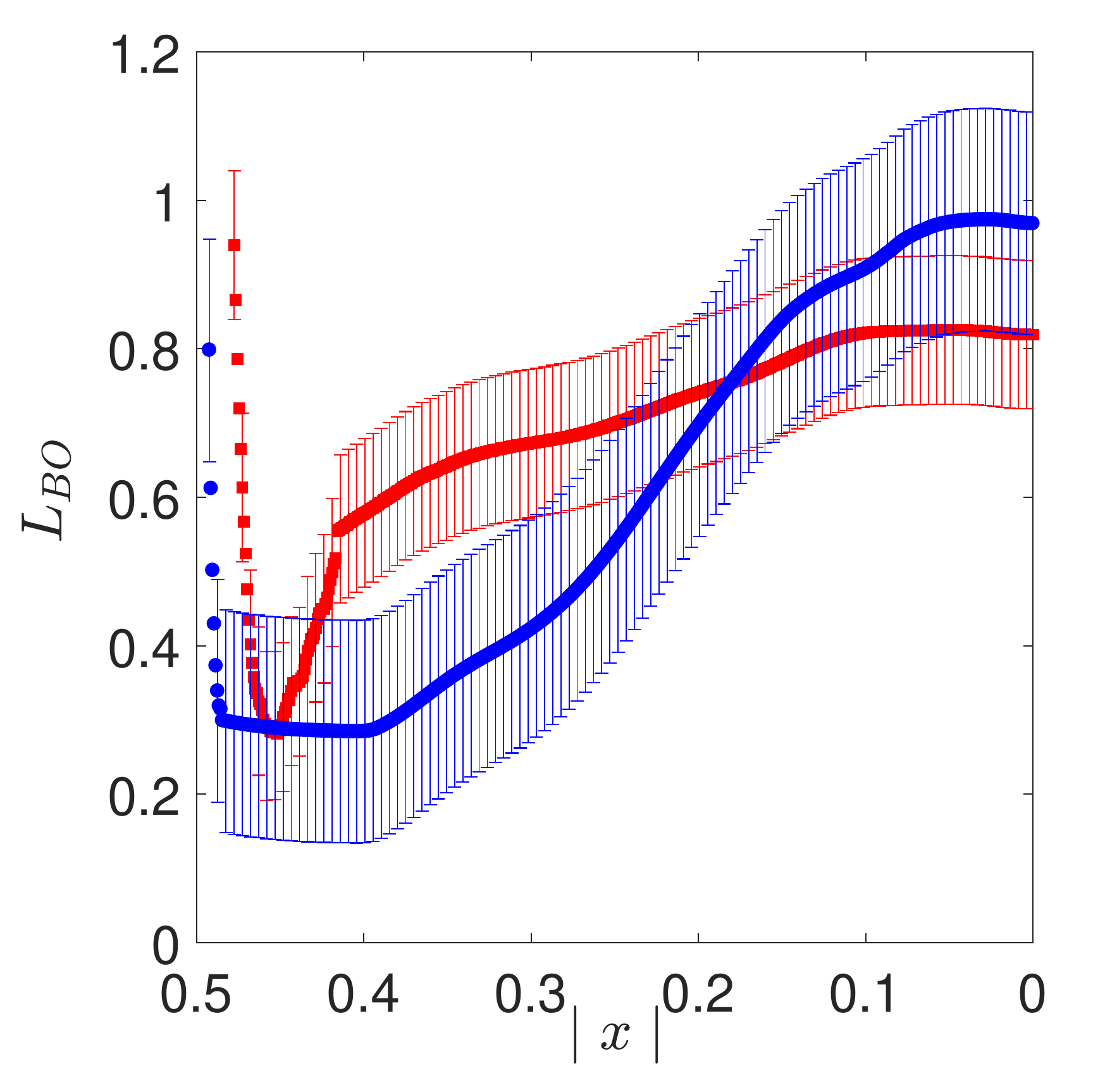}
\label{LOBprofile}}                  
\end{subfigure}
\begin{subfigure}[]
{\includegraphics[scale=0.33]{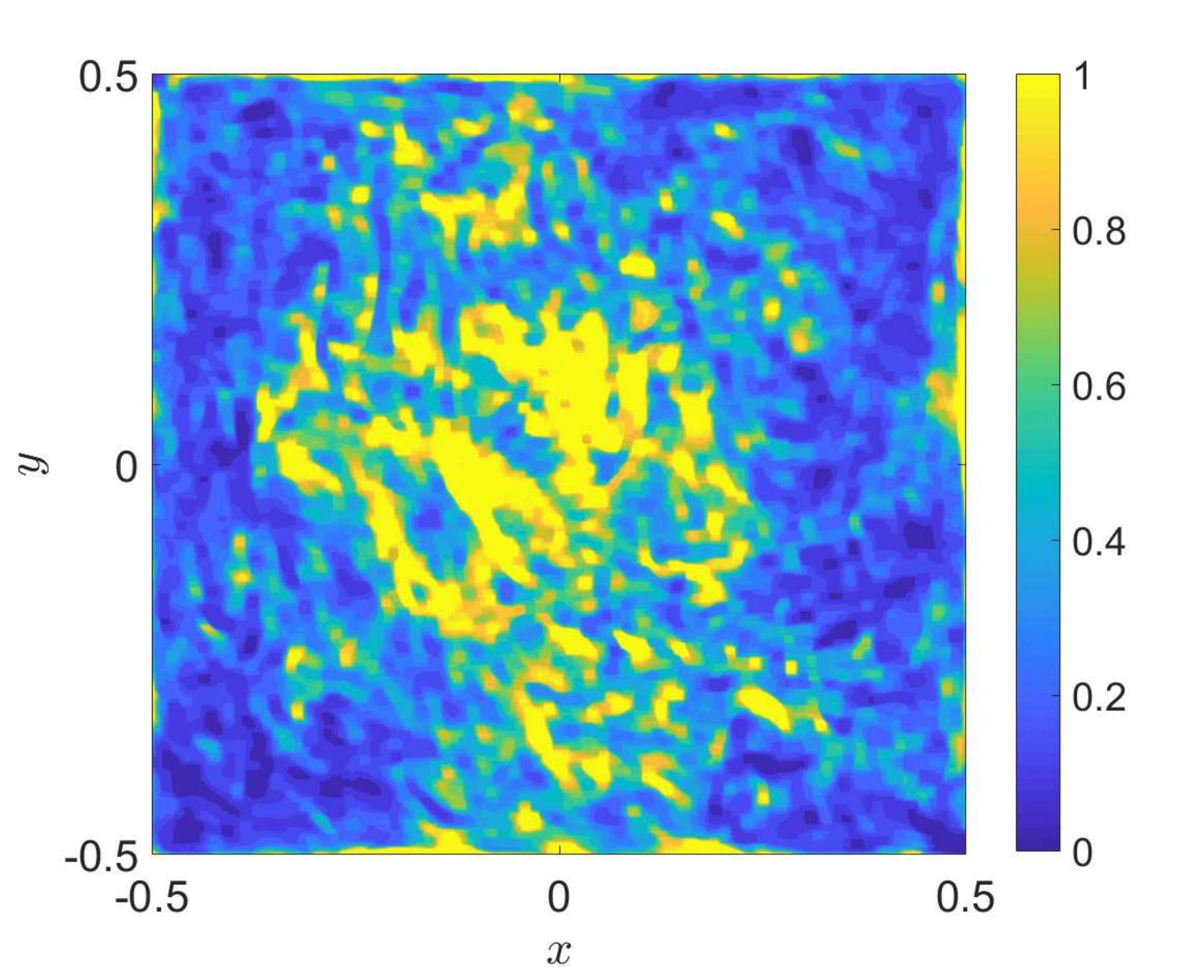}
\label{LOBcontourf}}                    
\end{subfigure}
\caption{\ref{LOBprofile}: Time-averaged profiles of the adimensional local Bolgiano-Oboukhov length scale ($L^{local}_{BO}$) in the horizontal direction ($x$), for $Ra = 10^7$ (red squares) and $Ra = 10^8$ (blue dots), with respective error bars.
{ The error bars are twice the standard deviation of $L^{local}_{BO}$, which is  globally computed for all $x$. }
 \ref{LOBcontourf}: Contour plot of the time-averaged adimensional local Bolgiano length scale on an horizontal section ($xy$) at half height of the cell {($z=0$)}, at $Ra = 10^8$.}
\label{Fig:3}
\end{center}
\end{figure}

To analyse this issue, in figure \ref{Fig:3} the  profiles of the $L_{BO}^{\textrm{local}}$ for the two $Ra$ numbers computed locally at the centre of the cube are shown. Both the entire horizontal plan and the length versus $x$ coordinate are reported. 
Due to the symmetry of the flow, we only plot half of the profile along the $x$ axis.
It turns out that large variations of $L_{BO}^{\textrm{local}}$ are experienced through the entire region, more evidently for $Ra=10^8$.
{ In particular, the length may be one order of magnitude shorter than the cell size \emph{locally}, over large span of the domain, at least for the higher $Ra$ case. This is highlighted in the map showed in figure \ref{Fig:3}b.}
From the pictures, it should yet be noted that quantities averaged only in time are not perfectly at convergence and some fluctuations are always present. Our results are hence to be considered qualitative but not necessarily quantitative.
The results are in any case very similar to those already presented in previous works~\citep{benzi1998heat,kaczorowski2013turbulent}.
Differences with respect to the more recent study appear in the statistical error bars, considering also that $Pr$ is slightly different, but it is known to have a huge impact.
Moreover, since we use a much higher resolution, some small differences in the evaluation of dissipation can be expected. 
Globally, we can consider that these two works validate each other, since they use different numerical approaches. 
Residual differences with respect to the older work by \cite{benzi1998heat}, are to be  explained  by the  changes in boundary conditions.
Our results therefore confirm also that the dynamics of the core region is significantly influenced by the boundary conditions.
Present results confirm the possibility of finding local Bolgiano scaling, separated from Kolmogorov-Oboukhov one, if local variations are properly computed.
\begin{figure}
\begin{center}
\begin{minipage}[c]{.35\textwidth}
\begin{subfigure}[]%
{\includegraphics[scale=0.25]{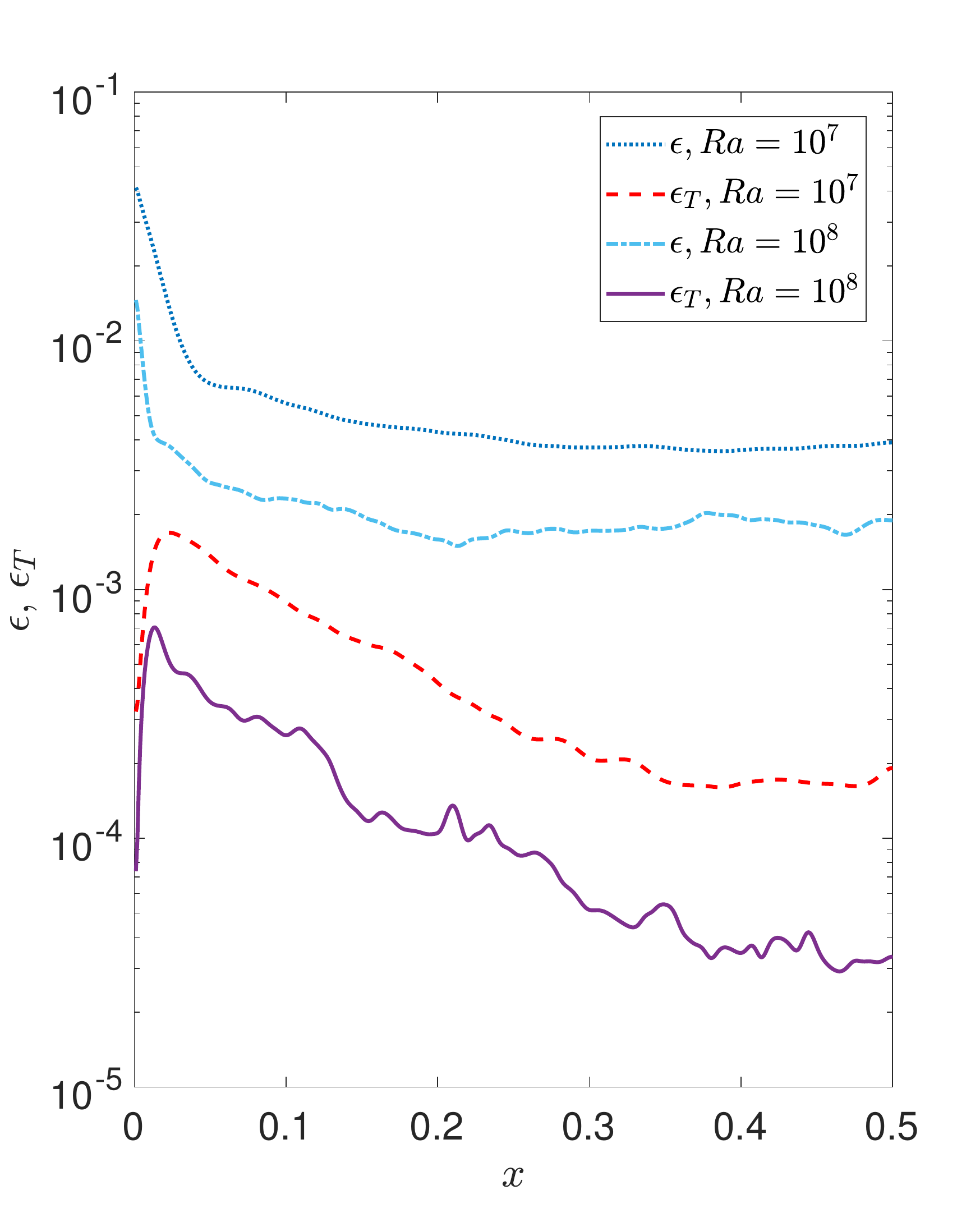}\label{epsilon}}                  
\end{subfigure}%
\end{minipage}%
\hspace{10mm}%
\begin{minipage}[c]{.35\textwidth}
\begin{subfigure}[]%
{\includegraphics[scale=0.21]{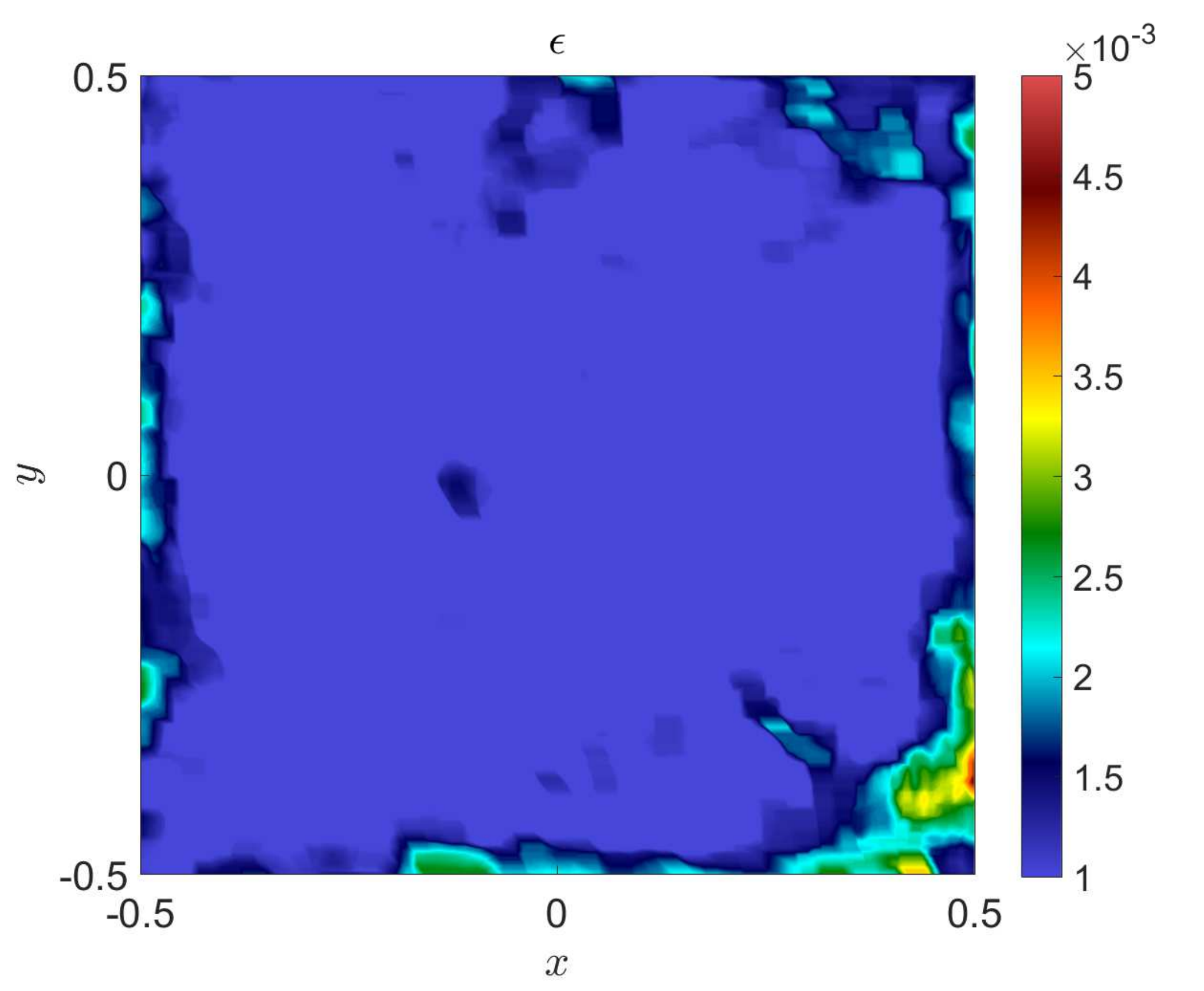}\label{epsiloncontourf}} 
\end{subfigure}\\%
\begin{subfigure}[]%
{\includegraphics[scale=0.2]{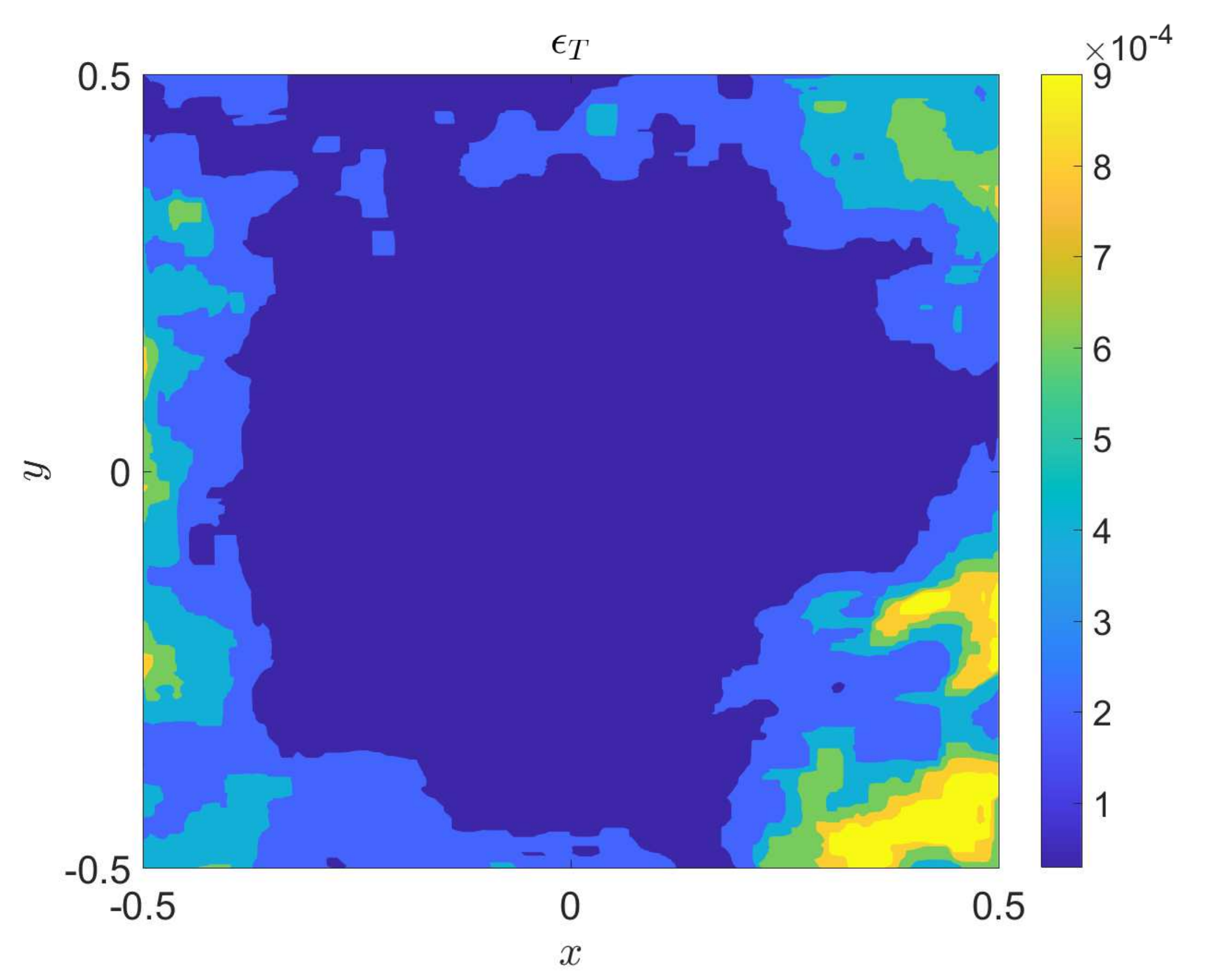}\label{epsilonT}}       
\end{subfigure}
\end{minipage}
\caption{\ref{epsilon}: Time-averaged profiles of the turbulent-kinetic-energy dissipation rate ($\epsilon$) and temperature variance dissipation rate ($\epsilon_T$) in the horizontal direction ($x$). Continuous line: $\epsilon_T$, at  $Ra = 10^8$;  dashed line: $\epsilon_T$, at $Ra = 10^7$;  dot-dashed line: $\epsilon$, at  $Ra = 10^8$; dotted line: $\epsilon$, at $Ra = 10^7$. \ref{epsiloncontourf}: Contourplot of the turbulent-kinetic-energy dissipation rate ($\epsilon$) values on an horizontal section ($xy$) at half height of the cell ($z=0.5$). \ref{epsilonT}: Contourplot of the temperature variance dissipation rate {($\epsilon_T$)} values on an horizontal section ($xy$) at half height of the cell ($z=0.5$).}\label{Fig:4}
\end{center}
\end{figure}

Since in the definition of $L_{BO}$ the energy and thermal variance dissipation are used, we show in figure \ref{Fig:4} the profiles of those functions, fig \ref{Fig:4}a, together with the surface contour for the $Ra=10^8$ case.
Once again, profiles obtained are similar to those shown in recent computations carried out in an analogous configuration~\citep{kaczorowski2013turbulent}.
The profiles obtained at two different $Ra$ numbers elucidate the dependence from the forcing parameter also of the Oboukhov-Bolgiano length. 
It should be clear also from these results that the direct measurement of this length based on its definition (\ref{eq:LOB}), is particularly delicate, because it is based on the ratio of powers of statistics of very small-scale observable, which widely fluctuate.
{ Furthermore, it should be emphasised that the definition (\ref{eq:LOB}) is physically sound but does not consider the characteristics of the flow, so that a  pre-factor (possibly non-universal) should be present, and there is no reason to be sure that this pre-factor is of order one~\citep{Mon_75}. 
 In the following section we shall estimate the Bolgiano-Oboukhov length looking at the scaling laws and we will compare the results. Specifically we will seek for the scale at which  a crossover in the scaling is encountered because of buoyancy.}

\subsection{Wavelet analysis}
\label{sec:wavelet}
Energy balances are computed on a horizontal section at half height of the Rayleigh-B\'enard cell. On this section, all the three components of the velocity field, and its derivatives on two dimensions, are taken into account.
Balances for a wide range of scales $\ell$ are calculated by using a Gaussian filtering function
$G_\ell(\textbf{r}) \sim e^{-\mid\textbf{r}\mid^2/2}$, where $\textbf{r} = \textbf{x}/\ell$ is a scale-dependent spatial coordinate.\\
In equations (\ref{DR}), (\ref{DRT}), (\ref{Dvisc}), (\ref{Dtherm}) and (\ref{exch}), the terms of the filtered kinetic and thermal energy balances defined in (\ref{filter-u}) and (\ref{filter-t}) are written explicitly as a function of the instantaneous velocity ($\textbf{u}$) and temperature ($\theta$) fields.

\begin{eqnarray}
\textrm{D}_{\ell} =  \frac{1}{4} \int_{}^{} d^dr (\nabla{G_\ell(\textbf{r})})\cdot [\textbf{u}(\textbf{x}+\textbf{r},t)-\textbf{u}(\textbf{x},t)]  [\textbf{u}(\textbf{x}+\textbf{r},t)-\textbf{u}(\textbf{x},t)]^2 
\label{DR}
\end{eqnarray}

\begin{eqnarray}
\textrm{D}^T_{\ell} =  \frac{1}{4} \int_{}^{} d^dr (\nabla{G_\ell(\textbf{r})})\cdot [\textbf{u}(\textbf{x}+\textbf{r},t)-\textbf{u}(\textbf{x},t)]  [\theta(\textbf{x}+\textbf{r},t)-\theta(\textbf{x},t)]^2 
\label{DRT}
\end{eqnarray}

\begin{eqnarray}
\textrm{D}^{\nu}_{\ell} = \frac{1}{\sqrt{Ra Pr}} \int_{}^{} d^dr (\nabla^2{G_\ell(\textbf{r})})\biggl [\textbf{u}(\textbf{x}+\textbf{r},t)\cdot\textbf{u}(\textbf{x},t)-\frac{\textbf{u}(\textbf{x}+\textbf{r},t)\cdot\textbf{u}(\textbf{x}+\textbf{r},t)}{2}\biggr ]  
\label{Dvisc}
\end{eqnarray}

\begin{eqnarray}
\textrm{D}^{\kappa}_{\ell} = \frac{1}{\sqrt{Ra Pr}} \int_{}^{} d^dr (\nabla^2{G_\ell(\textbf{r})})\biggl [\theta(\textbf{x}+\textbf{r},t)\cdot \theta(\textbf{x},t)-\frac{\theta^2(\textbf{x}+\textbf{r},t)}{2}\biggr]
\label{Dtherm}
\end{eqnarray}

\begin{eqnarray}
\textrm{D}^c_{\ell} = \frac{1}{2}\biggl [ \textbf{u}(\textbf{r},t)\cdot\frac{\textbf{g}}{\mid\textbf{g}\mid} \int_{}^{} d^dr ({G_\ell(\textbf{r})}) \theta(\textbf{x}+\textbf{r},t)+\theta(\textbf{r},t)\int_{}^{} d^dr ({G_\ell(\textbf{r})}) \textbf{u}(\textbf{x}+\textbf{r},t)\cdot\frac{\textbf{g}}{\mid\textbf{g}\mid} \biggr ]  \nonumber\\ 
 \label{exch}
\end{eqnarray} 

The convolution integrals in equations (\ref{DR}), (\ref{DRT}), (\ref{Dvisc}), (\ref{Dtherm}) and (\ref{exch}) can be computed very efficiently using continuous wavelets transforms ($WT$s), based on fast Fourier transforms. In this study, we use the 2D continuous wavelet MATLAB package provided by the toolbox YAWTB (http://sites.uclouvain.be/ispgroup/yawtb).\\
The terms $\textrm{D}_{\ell}$, $\textrm{D}^T_{\ell}$, $\textrm{D}^{\nu}_{\ell}$, $\textrm{D}^{\kappa}_{\ell}$, and $\textrm{D}^c_{\ell}$ are averaged over the time-length of each simulation (after reaching { statistically steady conditions}), and space in the bulk of the cell to get $\mathcal{D}_{\ell}$, $\mathcal{D}^T_{\ell}$, $\mathcal{D}^{\nu}_{\ell}$, $\mathcal{D}^{\kappa}_{\ell}$, and $\mathcal{D}^c_{\ell}$. 
{In particular, about $900$ independent snapshots of the flow have been used.
Statistical convergence of the budgets has been checked verifying that the results are basically unchanged using half of the data.
}

The bulk region is chosen by excluding 100 data points from each side, 
{ that is about 4 times the boundary layer thickness $\delta_{\theta}$ for the case at $Ra=10^7$, and about $7 \delta_{\theta}$ for $Ra=10^8$.
Indeed, in wavelet analysis on non-periodic flows, it is necessary to eliminate a border region to avoid spurious effects, and this choice of the bulk turns out to minimise the effects of the boundaries on the computation of the convolution integrals (\ref{DR})-(\ref{exch}). We have checked nevertheless that equivalent results are obtained using the half of the points and the double.
Concerning the boundary layer region,
always in order to avoid spurious effects from the border with the bulk, we have  chosen to keep about $8 \delta_{\theta}$ for the case at $Ra=10^7$, and about $15 \delta_{\theta}$ for $Ra=10^8$. We have checked that results are barely modified by taking a little larger or smaller regions.}


\subsection{Scaling laws}
\label{sec:scaling}

\begin{figure}
\begin{center}
{\includegraphics[width=1\textwidth]{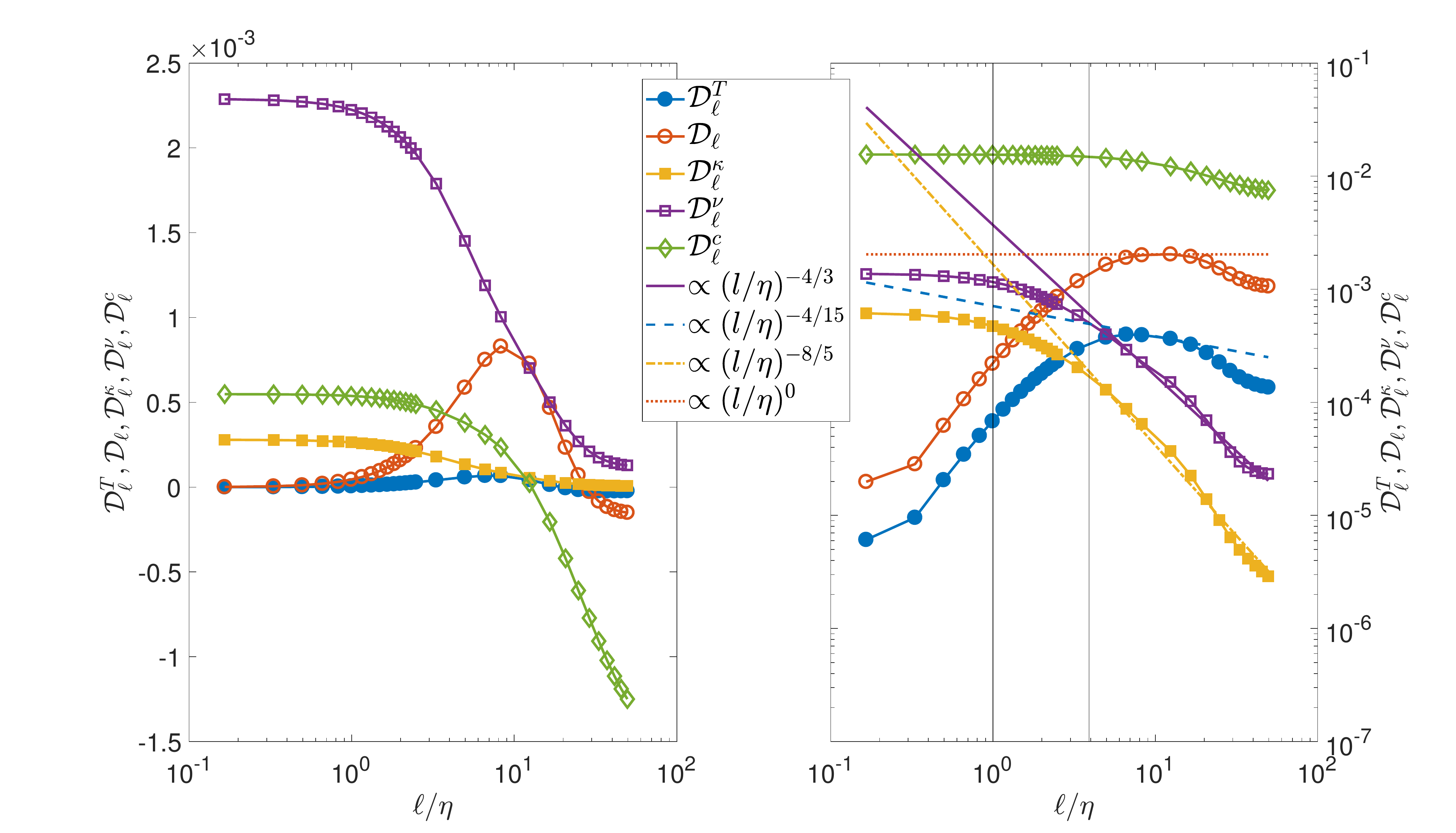}}
\caption{Time and space averaged energy balance terms, panel (a), and their time averaged spatial standard deviations, panel (b), as a function of scale $\ell$ over the Kolmogorov length scale $\eta$, { in the bulk region for $Ra = 10^7$}. Spatial averages and standard deviations were computed {in the bulk region}, on an horizontal slice at half height of the cube. The terms of the energy balance at scale $\ell$ are: $\mathcal{D}^T_{\ell}$: thermal energy term; $\mathcal{D}_{\ell}$: kinetic energy transfer term; $\mathcal{D}^{\kappa}_{\ell}$: thermal dissipation term; $\mathcal{D}^{\nu}_{\ell}$: viscous dissipation term; $\mathcal{D}^c_{\ell}$: exchange term between kinetic and thermal energy;
the two vertical black lines correspond to the Kolmogorov scale, $\eta$, and to the the scale where the Obouhkov-Bolgiano regime becomes visible in the scalings.}
\label{Fig:5}
\end{center}
\end{figure}

\begin{figure}
\begin{center}
{\includegraphics[width=1.0\textwidth]{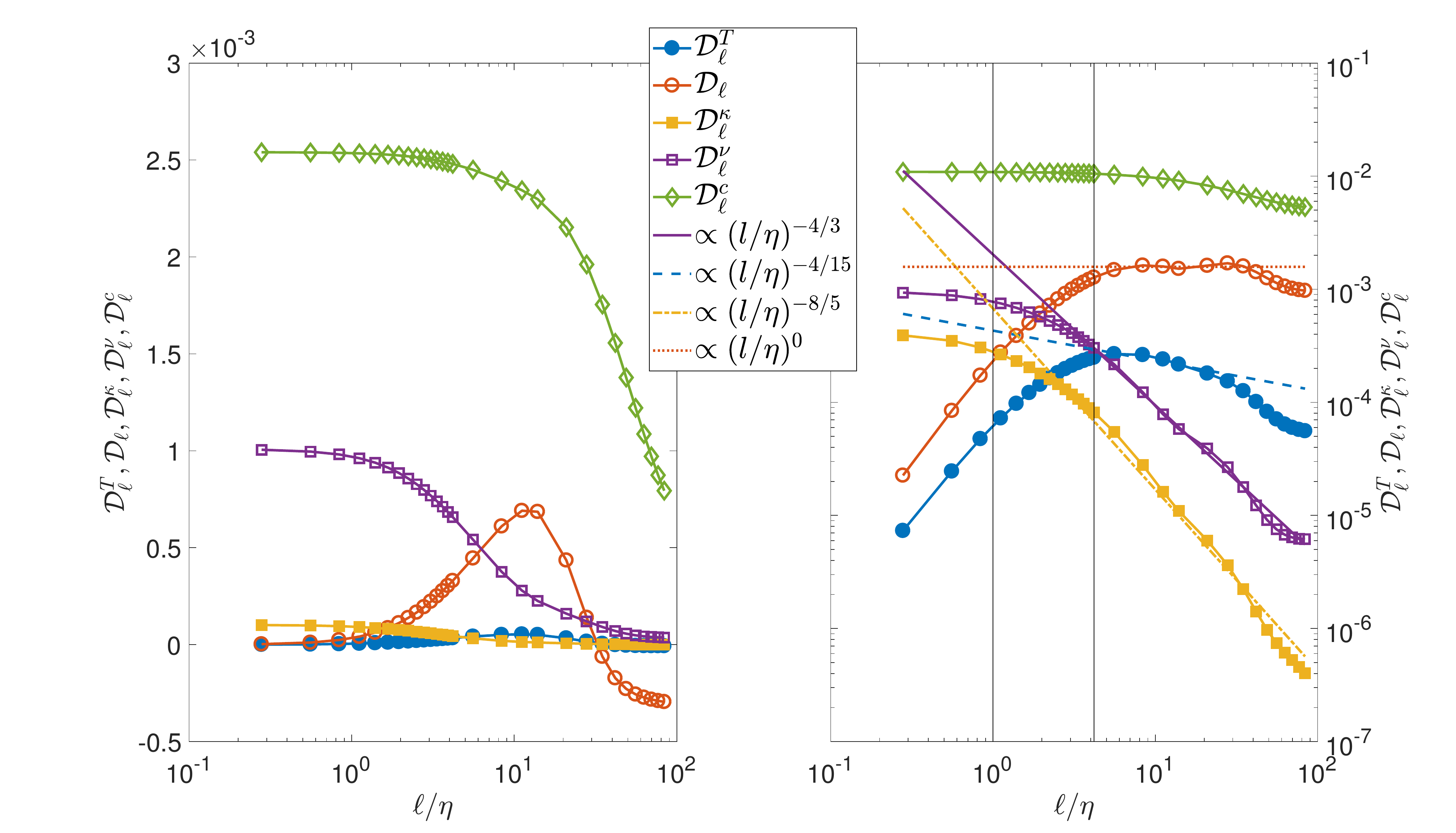}}
\caption{Time and space averaged energy balance terms, panel (a), and their time averaged spatial standard deviations, panel (b), as a function of scale $\ell$ over the Kolmogorov length scale $\eta$, {for $Ra = 10^8$}. Spatial averages and standard deviations were computed {in the bulk region}, on an horizontal slice at half height of the cube. The terms of the energy balance at scale $\ell$ are: $\mathcal{D}^T_{\ell}$: thermal energy term; $\mathcal{D}_{\ell}$: kinetic energy transfer term; $\mathcal{D}^{\kappa}_{\ell}$: thermal dissipation term; $\mathcal{D}^{\nu}_{\ell}$: viscous dissipation term; $\mathcal{D}^c_{\ell}$: exchange term between kinetic and thermal energy;
the two vertical black lines correspond to the Kolmogorov scale, $\eta$, and to the the scale where the Obouhkov-Bolgiano regime becomes visible in the scalings.}
\label{Fig:6}
\end{center}
\end{figure}

In figure \ref{Fig:5}-\ref{Fig:6}, we show the main results of the present work, that is the scaling behaviour of the mean and standard deviations of all terms of the equation (\ref{filter-u})-(\ref{filter-t}) once averaged over space and time for both $Ra$ numbers. 

Comparing figure \ref{Fig:5}a with  \ref{Fig:6}a, we see that the balance of terms depends strongly upon the Rayleigh number: at $Ra=10^7$, the viscous terms $\mathcal{D}^\nu_{\ell}$ is the largest at all scales, indicating that we are mainly in a dissipative regime. There is a small inertial interval around $\ell/\eta=10$, where the kinetic energy transfer term $\mathcal{D}_{\ell}$ and 
the thermal energy term $\mathcal{D}^T_{\ell}$ peak, indicating non-trivial turbulent behaviour. 
Indeed, the standard deviation (right panel) displays small inertial scaling range at this location. 
Moreover, the buoyancy term start being appreciable at $\ell \approx 5 \eta$ and becomes dominant at $\ell \approx 20 \eta$.
In contrast, for  $Ra=10^8$, the viscous term is dominant only up to $\ell \approx 5 \eta$, and  the coupling term $\mathcal{D}^c_{\ell}$ becomes dominant for $\ell \approx 25 \eta$, indicating a strongly convective regime. The kinetic energy transfer term $\mathcal{D}_{\ell}$ and 
the thermal energy term $\mathcal{D}^T_{\ell}$ still peak around $\ell/\eta=10$, with a wider  inertial scaling range for the standard deviation. 
{ Still concerning the exchange term, it is worth noting that at scales between the dissipative scale and the integral one this term may be negative. This is found to be particularly true in the regions not far from the walls.
However, it should be noted that the scaling analysis of the exchange term  $\mathcal{D}^c_{\ell}$ is rather inconclusive within present data indicating that boundary effects are important.}

{ As anticipated, standard deviation of the observables} display a much cleaner scaling than the mean quantities. We thus focus on them for the discussion of the scaling shape.
Looking at the scaling of fluctuations figs. \ref{Fig:5}b-\ref{Fig:6}b,
the dynamics appears qualitatively quasi-independent from $Ra$ number, at least in the present range, at variance with the scalings provided by global averages.
In practice, scalings at $Ra=10^8$ are clearer because of the absence of finite-$Re$ effect in this more turbulent regime. A  qualitative picture could however also be inferred from the lower $Ra$ number.
In any case, to be sure to avoid any viscous effect, we concentrate on the behaviour at $Ra=10^8$ in the following.

In the range $5 \eta  \lesssim \ell \lesssim 30 \eta$,  the buoyancy term is found to be greater than the non-linear thermal transfer but smaller than the kinetic energy one. 
This means that the Bolgiano-Oboukhov length can be physically estimated as $L_{BO}\sim 5 \div 10 \eta$, and that, starting from this point, Bolgiano scalings may be expected. 
{ This empirical estimate differs therefore from that obtained using the global definition (\ref{eq:LOB-norm}).}
{ In this range, the thermal dissipation term is consistent with ${\mathcal{D}}^{\kappa}\sim \ell^{-8/5}$, while the viscous one scales better like ${\mathcal{D}}_{\nu}\sim \ell^{-4/3}$. In both cases, the scaling appears to be robust and  extends over more than one decade.
Therefore, it turns out that in this range a Kolmogorov inertial scaling for the velocity but a Bolgiano-Oboukhov scaling for the temperature are found.
Finally the non-linear transfer term of the temperature equation is well reproduced by ${\mathcal{ D}^T}\sim \ell^{-4/5}$.
Below the Bolgiano-Oboukhov scale, for $\ell \lesssim 5 \eta$ scalings change clearly for all observable, but it is hard to extract the slope.
\begin{figure}
\begin{center}
{\includegraphics[width=0.39\textwidth]{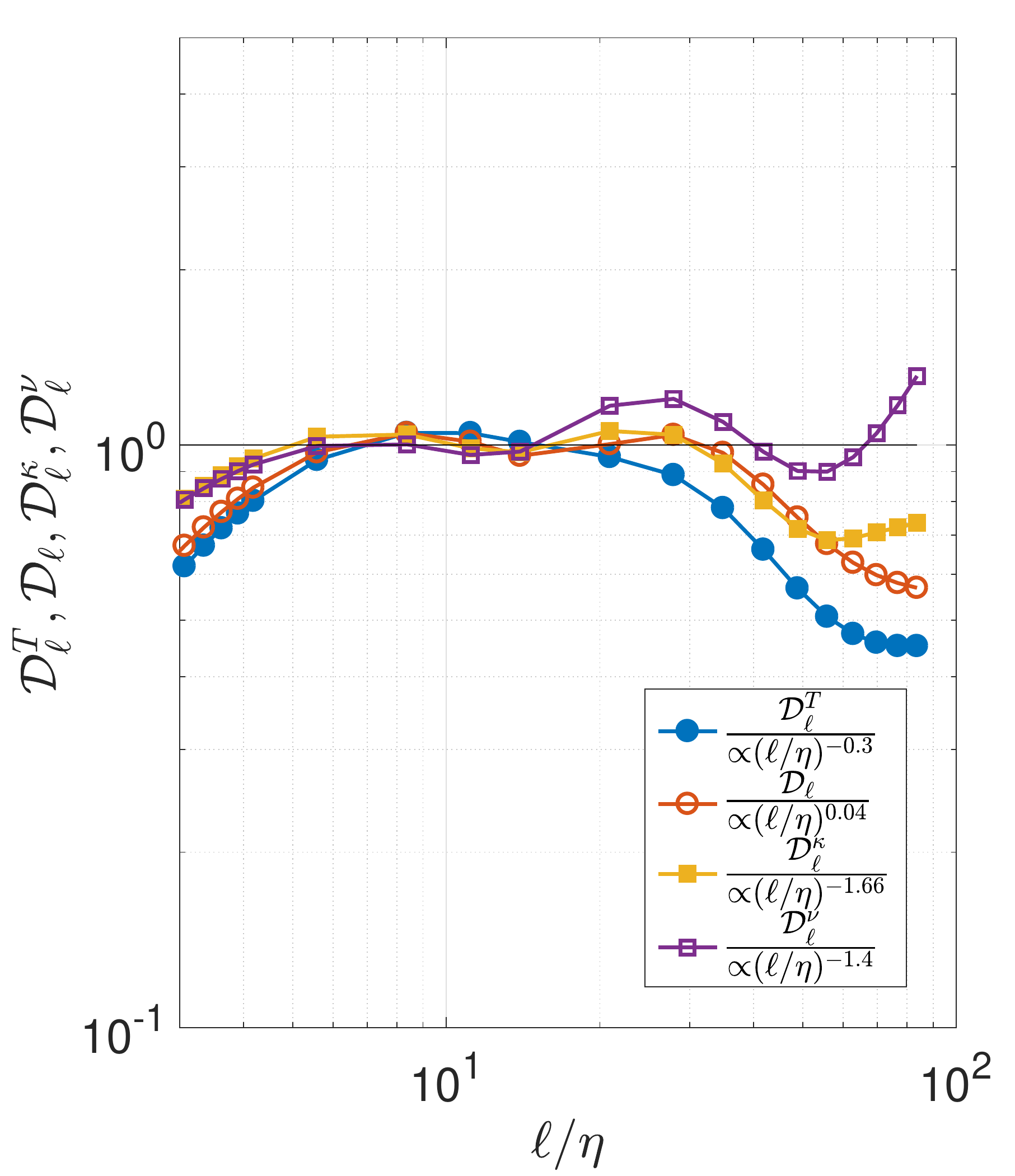}
\includegraphics[width=0.39\textwidth]{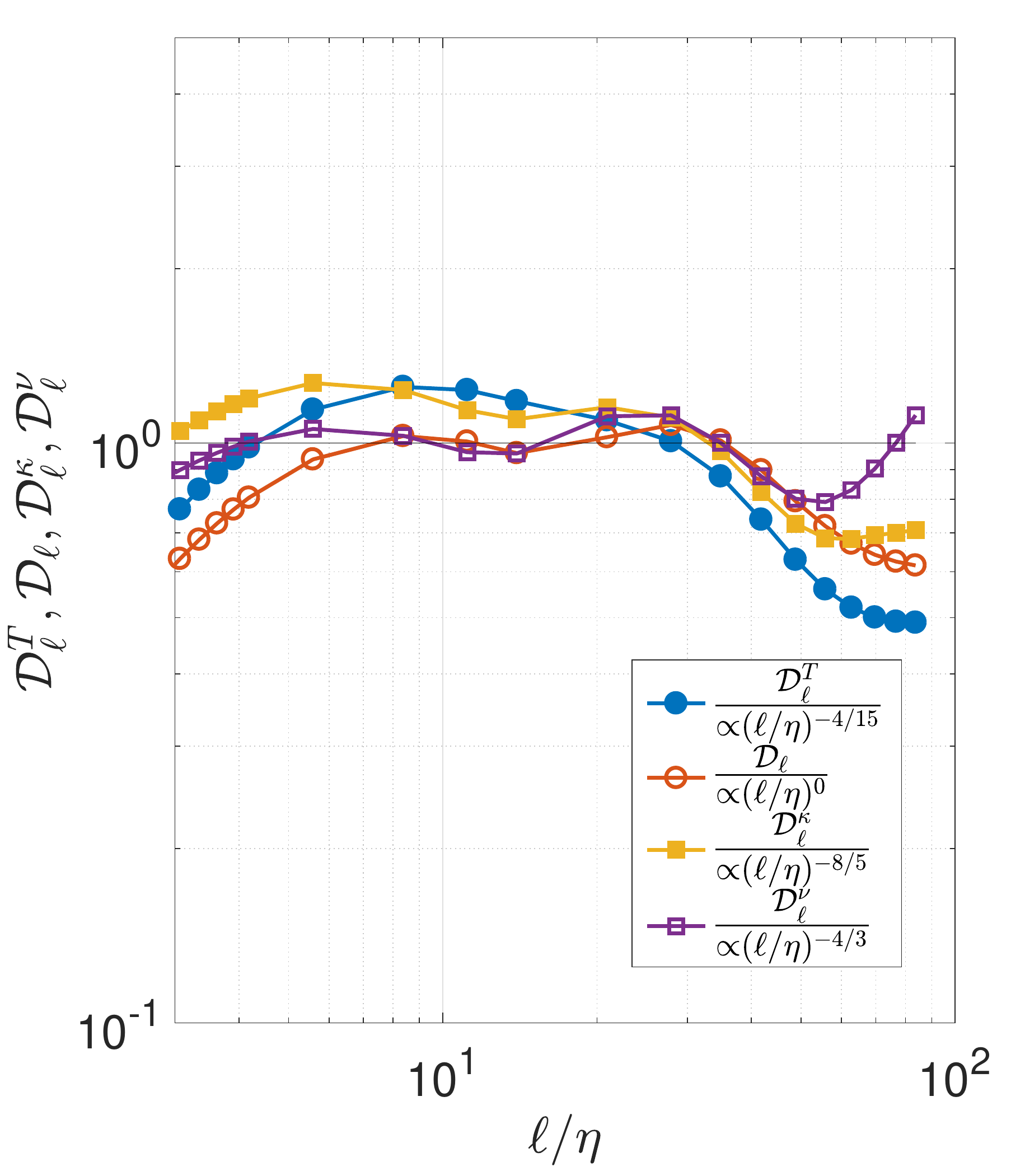}}
\caption{(a) Compensated plots of the budgets terms and the best-fitting curves. (b)  Compensated plots of the budgets terms and the conjectured scaling exponents related to similarity analysis.}
\label{Fig:6bis}
\end{center}
\end{figure}
{ As a conjecture we have used the typical exponents related to similarity analysis and the related curves are displayed in figures \ref{Fig:5}b-\ref{Fig:6}b.
To corroborate the picture, we have extracted the scaling exponents via the fitting of our data, and they are presented in table \ref{tab1} with the corresponding error for $Ra=10^8$.
We present also in figure \ref{Fig:6bis} the compensated plots both for the best fitting exponents and the conjectured ones.
It is seen that numerical exponents are all consistent with the similarity ones, which will be therefore used for the following discussion.}

We can now analyse the scalings found from figure \ref{Fig:6} in terms of scaling exponents within the general Kolmogorov-Onsager framework presented in section \ref{sec:KO}.
From $\ell  \gtrapprox 4 \eta$ a Bolgiano scaling is found for $\mathcal{D}^{\kappa}_{\ell}$, so that $h^T=1/5$. 
This is in line with the previous empirical estimate of $L_{BO}\sim 5\eta$, where we observed that the exchange term becomes more important than the thermal transfer term.
However, in the range $\eta < \ell \approx 30 \eta$,
the  velocity observables follow a Kolmogorov-like scaling with $\mathcal{D}_{\ell}\sim \ell^0~,\mathcal{D}^{\nu}_{\ell}\sim \ell^{-4/3}$, such that $h^u=1/3$.
Consistently, $\mathcal{D}^T_{\ell} \sim \ell^{h^u+2h^T-1}=\ell^{-4/15}$, as displayed in figure \ref{Fig:6} for $\ell \gtrapprox 5\div 10 \eta$.
Hence in this range the buoyancy term remains smaller than the non-linear inertial term of kinetic energy and  a Kolmogorov scaling is observed for the velocity.
In the range $\ell > 30 \eta$, buoyancy effect are dominant with respect to all other terms, and a pure Bolgiano-Oboukhov scaling should be present.  Boundary effects and lack of a sufficient number of scales make this conjecture speculative for the moment.
In the discussion section we suggest that anisotropy is responsible for the mixed scaling we obtain.
On the other side, in the range $\ell < L_{BO}\approx 5 \eta$,
all the scalings might be consistent with the Kolmogorov picture, even though we can only analyze a very small range of scales. In particular, $\mathcal{D}^T_{\ell} \sim\ell^{0}$, and 
$ \mathcal{D}^{\kappa}_{\ell} \sim \ell^{-4/3}$.
 
\begin{figure}
\begin{center}
{
\includegraphics[width=1\textwidth]{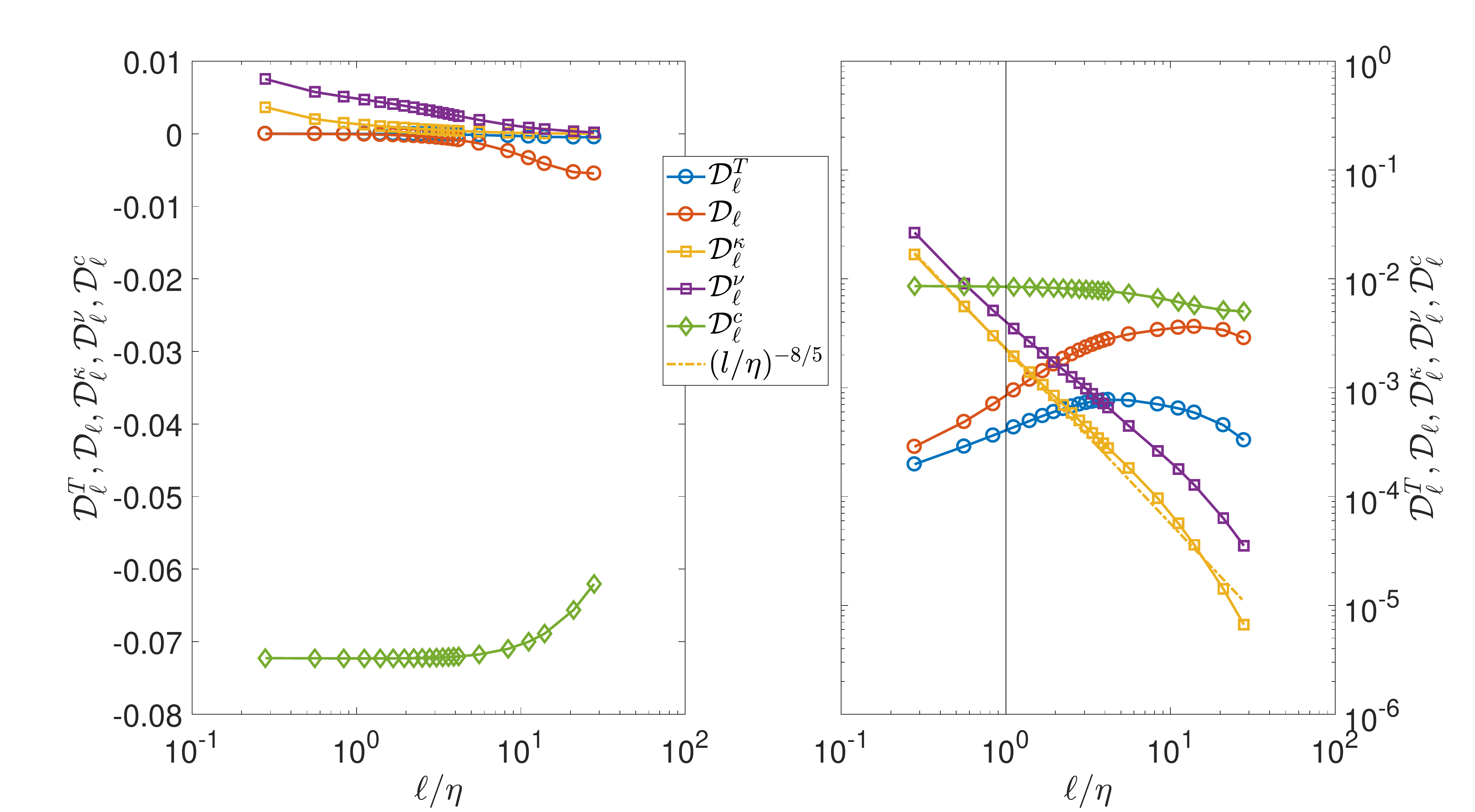}}
\caption{Time and space averaged energy balance terms, panel (a) and their time averaged spatial standard deviations, panel (b), as a function of scale $\ell$ over the Kolmogorov length scale $\eta$, {for $Ra = 10^8$}. Spatial averages and standard deviations were computed {in the boundary layer region}, on an horizontal slice at half height of the cube. The terms of the energy balance at scale $\ell$ are: $\mathcal{D}^T_{\ell}$: thermal energy term; $\mathcal{D}_{\ell}$: kinetic energy transfer term; $\mathcal{D}^{\kappa}_{\ell}$: thermal dissipation term; $\mathcal{D}^{\nu}_{\ell}$: viscous dissipation term; $\mathcal{D}^c_{\ell}$: exchange term between kinetic and thermal energy; 
the vertical black line correspond to the Kolmogorov scale, $\eta$.}
\label{Fig:7}
\end{center}
\end{figure}
It is interesting to look also at statistics in the region near to the walls, that is in the boundary layer.
Indeed as displayed in Fig. \ref{Fig:3}, we expect that the Bolgiano-Oboukhov has a minimum in the boundary-layer region at few points very close to the walls, and therefore buoyancy effects should start being dominant even at a smaller scale.
Yet, it is important to note here that with respect to the results obtained in the core region, the boundary layer statistics should be handled with much care and in no case considered as conclusive, for several reasons.
To get these statistics we consider { about $15 \delta_{\theta}$, which means} only the first $200$ points near to each boundary, so that the number of scales available is small and no clear scale  separation can be expected.  
Furthermore, these regions are strongly non-homogeneous and impacted by viscous effects, inducing without any doubt spurious effects on statistics, so that the significancy of the results must not be considered certain.}
In figure \ref{Fig:7}, we show the results obtained for the different observables at $Ra=10^8$.
Even with the caveat about the applicability of the theory near the boundaries, the results are interesting. 
In this region a Bolgiano-Oboukhov scaling seems to be found for both temperature and velocity in the range $\eta \lesssim \ell \lesssim 10 \eta$, in particular  
{ scalings are consistent with }
$\mathcal{D}_{\ell}\sim \ell^{4/5}~, 
~\mathcal{D}^T_{\ell} \sim \ell^{0}~, 
 \mathcal{D}^{\kappa}_{\ell} \sim \ell^{-8/5}$.
Yet the viscous term appears to have a scaling ${\mathcal{D}}_{\nu}\sim \ell^{-1}$. 
It is  therefore a little steeper than what expected in the BO range.
These findings confirm the importance of local properties of the fields and the necessity to accurately disentangle them from more global effects.
Since near the boundaries $L_{BO}$ becomes small, the Bolgiano scalings are more effective for both velocity and temperature.
The slightly inconsistent  behaviour of the velocity dissipation term is thought to be related to finite $Re$-effects, since near the boundaries { the local Re number is not large} and the similarity arguments \emph{\`a la Kolmogorov} are not expected to hold.

\section{Discussion and Conclusions}
We have carried out a very high-resolved DNS analysis of the small-scale properties of turbulent Rayleigh-B\'enard convection in a cubic cell at $Pr=1$. The  unusually accurate resolution allows to go well below the Kolmogorov length.
We have used two simulations at $Ra=10^7$ and $Ra=10^8$.
{ It has been longly known that there is a transition between a chaotic to a fully developed state  around these values of $Ra$~\citep{siggia1994high}.
Although it is now widely accepted that the transition is not sharp as initially guessed~\citep{castaing1989scaling}
it is interesting to capture possible signatures of a transition in scaling laws.
Our main goal was to apply a new approach based on the weak formulation of the mathematical problem, and to extract in this way scaling exponents.}

{ Previous experiments and numerical simulations have shown that Bolgiano-Oboukhov scaling should be easier to be observed at higher $Pr$ number~\citep{kaczorowski2013turbulent}. However, the dependence on the Pr number turned out to be moderate, and given that we need a very high resolution to properly compute local scaling at small scales,  numerical simulations respecting the level of accuracy we have required in the present work would be unfeasible. Considering also that our main goal here was to show how new insights may come from the weak approach to turbulent problems, that explains why we have simulated a flow with $Pr=1$.
The question of investigating some local properties at higher Pr remains nevertheless relevant in many respects. For this reason, we have planned the computation of Rayleigh-B\'enard convection at $Pr=7$, yet at more moderate $Ra$.
We hope to report the findings in a future work.   }

As already indicated in previous numerical works, we have found that the local Bolgiano-Oboukhov length is a strongly varying function of the position, with the maximum at walls and a minimum approximately at the end of the boundary layer. The complex behavior of the length is due to its dependence on the thermal and kinetic dissipation. 
The global Bolgiano-Oboukhov length is found more or less the same for the two $Ra$ and of the order of the fluid layer height.
Globally speaking, the results concerning the statistical observables of the flow are in good agreement with previous results. Since our simulations have been conducted with a different method than in the other studies, that is an indication of robustness that corroborate the findings.

From the theoretical point of view, we have presented the derivation of the weak-formulation or coarse-grained version of the Boussinesq equations for turbulent convection which allow a smoothed treatment of instantaneous fluctuations.
The resulting set of equations have been averaged in the present work to get the generalization of the K\`arm\`an-Howarth-Monin and Yaglom equation for the fully non-homogeneous problem. 
In our geometry, these equations are the analogous to those obtained in recent work directly from the Boussinesq equations~\citep{rincon2006anisotropy} and permit a clear scale-by-scale analysis of the turbulence cascade in the physical space.
{ This original approach is useful to obtain scaling behaviour in a less noisy manner with respect 
to more standard statistical procedures. Moreover, when applied to the fluctuating equations, it allows to get information on the probability distribution of fluxes,
which is crucial to characterise extreme events and intermittency. Therefore, the approach presented in this work should be valuable to  get new insights also in convective turbulence}.

Using this filtering approach, we have analyzed the scalings characterizing the kinetic energy and temperature variance cascade. 
The different scalings we found are reported in Table \ref{tab:scaling}. 
{ We have not found evidence of a standard Bolgiano-Oboukhov scaling, which would mean $h_u=3/5~,h_T=1/5$.
Instead, our numerical experiment points out that only the temperature follows this Bolgiano scaling, so that  buoyancy effects are found to be dominant on the temperature variance budget at small scales.
Yet 
the velocity  follows the
Kolmogorov 41 scaling $h_u=1/3$, at least in the available range of scales.\

Such peculiar behaviour can be explained by removing the isotropic condition, and consider that the horizontal velocity increments and the vertical velocity increment scale with a different 
exponent, respectively $h_u^H$ and $h_u^V$. In such a case, it is easy to see that the scaling exponent  ${\mathcal{ D}}$,  ${\mathcal{ D}}^T$, 
${\mathcal{ D}}^\nu$,  ${\mathcal{ D}}^\kappa$
will be respectively $\min(3h_u^H-1,3h_u^V-1)$, $\min(h_u^H+2h_T,h_u^V+2h_T)$, $\min(2h_u^H-2,2h_u^V-2)$, $2h_T-2$. If we take $h_u^H=1/3$,  $h_u^V=3/5$ and $h_T=1/5$, we thus get 
the theoretical results of Table \ref{tab:scaling}.

Moreover, the present results are compatible with the previous studys by \cite{kunnen2008numerical,kaczorowski2013turbulent}. Although the scaling were extracted on very few points, temperature structure functions indicate a Bolgiano-Oboukhov scaling whereas the velocity structure functions were less well defined. More importantly, only the axial functions showed some hint of   Bolgiano-Oboukhov scaling, while the horizontal ones had no conclusive scaling. That points to a possible anisotropic effect which may affect only velocity. Moreover, in the numerical study by~\cite{camussi2004temporal} a BO scaling for the temperature and a K41 for the velocity were also found.
In experiments, while evidence of Bolgiano scaling on the temperature are available since some time \citep{wu1990frequency,cioni1995temperature,ashkenazi1999spectra}, velocity scaling is more elusive and the effect of anisotropy has been also reported~\citep{ching2004velocity,sun2006cascades,ching2007scaling}. 

Then the presence of lateral walls is found to be key in the possible changing of scaling in the core of the flow.
Indeed because of walls: (a) the local Bolgiano-Oboukhov length experiences large variability and notably may be ten times less than the global one that is of the order of the cell length. This explains why in horizontal homogeneous simulations the  Bolgiano-Oboukhov range should not be observed at least in the core of the flow~\citep{lohse2010small,verma2017phenomenology}. (b) Since the flow is non-homogeneous, the budget equation for kinetic energy and temperature variance are complex, and transport terms play locally a role as recently emphasised in a scale-by-scale analysis using another approach~\citep{togni2015physical}. 
In particular, it is found that at variance with the homogeneous case the coupling term may be locally negative at small scales, and in particular in the vicinity of boundary layer, so that kinetic energy is converted in potential one. Our numerical evidence hence confirms a previous theoretical analysis~\citep{l1991spectra,l1992conservation}. Instead, in the homogeneous case it is found the contrary which may lead to the impossibility to observe a Bolgiano-Oboukhov scaling~\citep{verma2017phenomenology}.
(c) The contribution of the buoyancy coupling term is found to be important at all scales as reported previously~\citep{togni2015physical}, however its relative importance with respect to other terms differs. Notably, it is found to be dominant in the temperature budget at almost all scales, whereas the inertial term of the kinetic energy budget is is the most important term at small scales. 

From the above considerations, we can draw the following picture:
the presence of walls makes the local Bolgiano length small, and buyoancy is effective on the budget of temperature variance over a wide range of scales. That allows the emergence of a BO scaling on the temperature and on the vertical velocity component in the whole core region. On the other hand, in the bulk region the nonlinear inertial term remains much greater than the buyoancy one and starts decreasing only at scales too large to allow the identification of a possible BO scaling. As for the horizontal velocity components,they are less affected by buoyancy, hence the KO41 scaling observed. In the boundary layer region, the local Bolgiano length may be much smaller than the global one and the non-homogeneous character of the region makes the redistribution among velocity components important. 
The local approach used here indicates indeed a possible BO scaling for the temperature and velocity, but some discrepancy in the viscous term that is attributed to a finite-Re correction. 
It is worth emphasising however that the small number of points available in the boundary layer and the difficulties inherent to such a non-homogeneous region make our reasoning not at all definitive, and ask for a deeper analysis of the issue.
In particular, the scaling of the velocity in this region is also compatible with $h_u=2/3$ which is typical of shear flows~\citep{biferale2005anisotropy}. 
}
As far as it concerns the effect of Rayleigh number, comparing two set of results obtained at different $Ra$, we have also shown that some differences are related to the transition from { a  chaotic ($Ra=10^7$) to a more turbulent regime ($Ra=10^8$)}. In particular, the viscous terms in the energy and temperature budget are important at all scales at $Ra=10^7$, and not only in the boundary layer. That indicates that much of the transport is always due to viscous diffusion, whereas it becomes negligible  at $Ra=10^8$ at least in the core of the flow.

In the present paper, we have focused only on average quantities, performing the average over selected portions of the domains. The interest of our formulation, however, is that it also provides expression of {\sl local} transfer quantities. It would be interesting to connect those local energy transfer to possible intermittency of the convective fluids. 
We leave that for future work.

\section*{Acknowledgments.}  
This work has been supported by  the ANR EXPLOIT,  grant agreement no. ANR-16-CE06-0006-01.
We acknowledge the CINECA for the availability of high performance computing resources and support. The computations on Marconi were supported by a PRACE grant.

\clearpage
\bibliographystyle{jfm}
\bibliography{paper}

\end{document}